\documentclass[%
	reprint,
superscriptaddress,
 amsmath,amssymb,
 aps,
prab,
]{revtex4-1}

\usepackage{graphicx}
\usepackage{dcolumn}
\usepackage{bm}
\usepackage{units}
\usepackage{siunitx}
\usepackage[colorinlistoftodos]{todonotes}

\usepackage{hyperref}
\usepackage{xcolor}
\hypersetup{
    colorlinks,
    linkcolor={red!50!black},
    citecolor={blue!50!black},
    urlcolor={blue!80!black}
}

\renewcommand{\d}[2]{\frac{\mbox{d} #1}{\mbox{d} #2}} 
\usepackage{mathtools}

\DeclarePairedDelimiter\floor{\lfloor}{\rfloor}

\begin{document}

\preprint{APS/123-QED}

\title{Compton recoil effects in staging of laser wakefield accelerators}

\newcommand{\JAI}{The John Adams Institute for Accelerator Science, Imperial College London, London, SW7 2AZ, UK}

\author{M.\,J.\,V.\,Streeter}
\affiliation{\JAI}
\author{Z.\,Najmudin}
\affiliation{\JAI}

\date{\today}

\begin{abstract}

Laser plasma accelerators capable of generating \unit[$>10$]{GeV} electron beams may require plasma mirrors to remove undepleted laser energy at the end of each accelerator stage. 
Near the plasma mirror surface, the electron bunch can interact with the reflected light, resulting in inverse Compton scattering.
For realistic conditions, we show that a significant fraction of electrons emit one or more photons, increasing the energy spread of the electron bunch. 
We provide an analytical expression for calculating this effect, and use it to estimate the minimum drift space required before the plasma mirror to meet given energy spread specifications.
Mitigation strategies, necessary to achieve sub-percent energy spread in multi-GeV laser wakefield electron sources, are proposed and explored.
\end{abstract}

\maketitle

\section{Introduction}
Laser wakefield accelerators (LWFA) \cite{Tajima1979PRL} are capable of producing accelerating gradients of \unit[$\sim100$]{GeV/m} \cite{Mangles2004N,Faure2004N,Geddes2004N}.
The energy an electron can gain using a single laser pulse is fundamentally limited by dephasing of the electrons relative to the accelerating field and by depletion of the driving laser pulse.
The maximum electron energy before reaching these limitations can be increased by using a lower plasma density.
However, this requires higher energy laser pulses, increasing the cost and size of the accelerator.
An attractive alternative is to couple together multiple acceleration stages \cite{Leemans2009PT,Steinke2016N}, each of higher accelerating gradient but shorter length, as is common in large-scale conventional accelerators.
This would enable plasma based accelerators to achieve electron energies required for high-energy physics applications, such as an electron-proton collider using an electron beam at \unit[50]{GeV} \cite{Wing2019RS}, using currently available laser powers.
However, there are many challenges for staged LWFA to reach this energy level with the requisite high bunch quality.

For staged LWFA, it is necessary to couple a new laser pulse onto the acceleration axis at the start of each stage \cite{Leemans2009PT}.
A significant fraction of the laser energy remains at the exit of a LWFA \cite{Streeter2018PRL}.
Therefore, a compact method is also required to extract this undepleted laser energy to avoid damage to subsequent beam optics, diagnostics and other devices.
One method for coupling both incoming and outgoing laser beams, is to use plasma mirrors \cite{Ziener2003JAP}.
The plasma mirror can be in the form of a thin tape which is instantaneously turned into a high-density and hence highly-reflective plasma by incidence of the high power laser beam. Since part of the tape is vapourised each time, it must be translated to a new position between shots.
The tape must be thin in order to minimise detrimental effects on the electron beam due to scattering, but also strong enough to survive mechanically.

\begin{figure}[htb] 
   \centering
   \includegraphics[width=8.5cm]{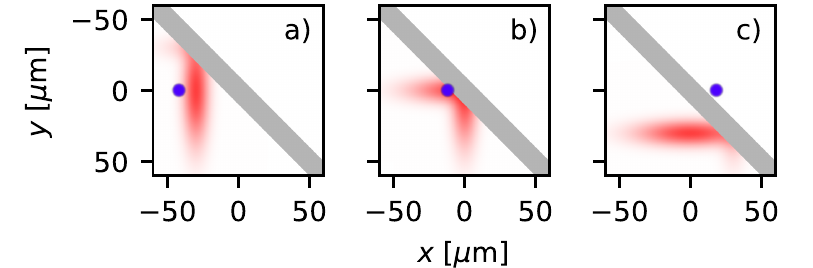} 
   \caption{Schematic of the interaction of an LWFA electron bunch (blue) with the laser pulse (red) reflected from a thin foil (grey):
   a) Electron beam trails laser pulse  propagating in the positive $x$ direction.
   b) Laser pulse is reflected from foil and interacts with electron beam.
   c) Electron beam passes through foil and is separated from laser field.}
   \label{fig:ICS_picCore}
\end{figure}

One consequence of using plasma mirrors, which has not previously been discussed, is the effect of inverse Compton scattering (ICS) on the electon beam.
Due to the physical size of the laser pulse in a LWFA, and the distance at which the electron bunch trails the laser pulse, it is possible for the electron beam to enter the field of the reflected laser before it crosses the boundary of the plasma mirror, as shown in fig.\,\ref {fig:ICS_picCore}.
In this region, the electrons oscillate in the laser field and can emit radiation.
This mechanism has previously been demonstrated with a zero degrees plasma mirror as an all-optical source of gamma-rays \cite{TaPhuoc2012NP}.
For high electron energies, the emitted radiation can take a significant fraction of the electron's energy.
In strong laser fields, this leads to the `radiation reaction' problem which has been the subject of recent experimental \cite{Cole2018PRX,Poder2018PRX} and theoretical \cite{Blackburn2014PRL,Vranic2016NJP,Ridgers2017JPP} study.
If the energy loss is significant and is experienced by an appreciable proportion of electrons, it will adversely affect the electron bunch quality, particularly the energy spread.
We have conducted the following analysis in order to determine the severity of the issue, and how it may be mitigated.

\section{LWFA design}
An LWFA consists of an electron plasma wave driven by the ponderomotive force of an intense laser pulse.
The phase velocity of this plasma wave is determined by the group velocity of the laser in the plasma; $v_g = c\sqrt{1-(\omega_p/\omega_0)^2}$, where $\omega_p = \sqrt{n_e e^2/\epsilon_0 m_e}$ is the plasma frequency for a plasma of density $n_e$, and $\omega_0$ is the laser frequency.
For low plasma density ($\omega_p \ll \omega_0$), the Lorentz factor associated with the plasma wave is therefore $\gamma_{\phi} \approx {\omega_0}/\omega_p$.
The field structure of the plasma wave provides focusing and accelerating regions moving at close to the speed of light and is thus ideally suited to accelerate electrons.
However, a relativistic electron beam injected at the rear of the accelerating and focusing region will advance relative to the plasma wave leading to eventual dephasing.

The properties of the wakefield are determined by the normalised vector potential $a$ of the laser pulse, which has a maximum value $a_0= e E_0/(m_e \omega_0 c)$.  
In the linear regime ($a_0<1$), the (maximum) electron energy gain at dephasing is $\gamma_{\max}m_e c^2= 2({\omega_0}/\omega_p)^2\, m_e c^2$ after a dephasing length $k_p L_{\phi} = (\omega_0/\omega_p)^2$ \cite{Lu2007PRSTAB}, where $k_p = \omega_p/c$.
For a strong drive laser ($a_0 \gg1$), the plasma wave is highly non-linear.
The plasma density becomes cavitated, as a large fraction of electrons are expelled from the regions of highest laser intensity, and so the focusing and accelerating fields increase and cover a larger fraction of the plasma wave period \cite{Lu2006PRL}.
non-linear evolution of the laser pulse envelope becomes more important, and pulse front etching leads to a reduction in phase velocity \cite{Decker1996POP}.
As a result, the maximum electron energy is $\gamma_{\mathrm{max}} m_e c^2 = \frac{2}{3} a_0 ( \omega_0/\omega_p)^2\, m_e c^2$ which occurs after a dephasing length $k_p L_{\phi} = \frac{4}{3}\, \sqrt{a_0} (\omega_0/\omega_p)^2$ \cite{Lu2007PRSTAB}.

Both the highly non-linear `blowout' ($a_0 \gg1$) and the quasi-linear regimes ($a_0 \gtrsim 1$) have been used to accelerate electrons to multi-GeV energies.
The acceleration length can be extended to the dephasing length if the laser power exceeds the threshold for relativistic self-guiding \unit[$P_{\mathrm{crit}}=17 (\omega_0/\omega_p)^2$]{GW}.
Relativistic self-guiding has been demonstrated over $>100$ Rayleigh ranges \cite{PoderPPCF2018} and has been used to generate electron energies of \unit[$>2$]{GeV} \cite{Wang2013NC,Kim2013PRL}.
For laser power $P\leq P_{\mathrm{crit}}$, an external guiding structure is required \cite{Butler2002PRL,Shalloo2018PRE}.
External guiding allows for LWFAs to operate at lower plasma densities, generating electron energies up to \unit[8]{GeV} \cite{Gonsalves2019PRL} in a single stage.

In order to reach electron energies $\gamma_{\rm max}m_e c^2 \gg \unit[10]{GeV}$, multiple LWFAs must be used to provide successive acceleration stages for the electron bunch.
In a multi-stage LWFA, a plasma mirror would be required between each stage to extract the residual laser energy from one stage before coupling in the laser pulse for the next one.
At the plasma exit, the laser pulse transverse spot size is determined by the guiding method.
For self-guiding in the `blowout' regime, the laser transverse size is approximately the plasma bubble radius $k_p R_b = 2\sqrt{a_0}$ \cite{Lu2007PRSTAB}.
For external guiding, the matched spot size is set by the transverse plasma density profile. 
In the design by Schroeder \emph{et al.}~\cite{Schroeder2010PRSTAB} a focal spot radius of \unit[70]{$\mu$m} is given for an externally guided \unit[$10$]{GeV} stage.

An electron bunch accelerated in the first plasma period will trail the laser pulse by approximately half of the plasma wavelength.
Then, if the laser is reflected by a \ang{45} plasma mirror, the electron beam will pass through part of the laser field and Inverse Compton scattering will occur.
The scattered photons can take a significant fraction of the initial electron energy and are emitted at small angles to the electron direction of motion. 
Electrons involved in the scattering interaction lose energy as a result, causing an increase in the energy spread of the electrons beam.
Thus, this effect must be carefully managed for LWFAs to maintain narrow energy spread.

\section{ICS photon energy and cross-section} 
\begin{figure}[hpbt] 
   \centering
   \includegraphics[width=8.5cm]{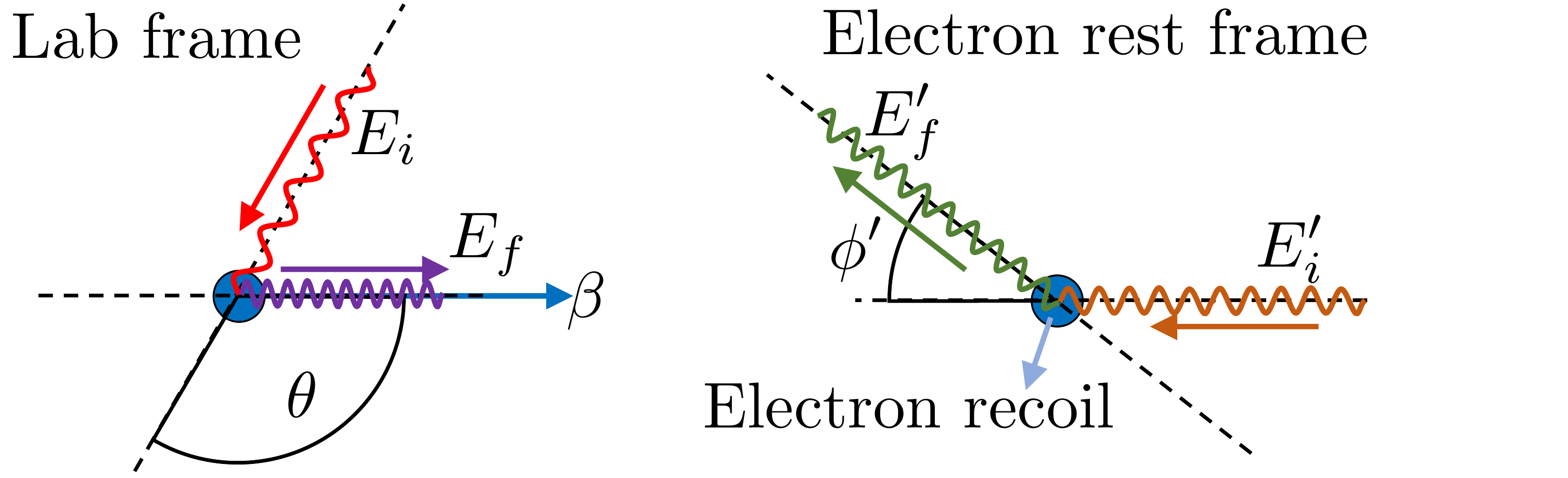} 
   \caption{Illustrations of the ICS geometry in the lab- and electron rest-frames.
   With the assumptions $\gamma\gg1$ and $\theta \gg 1/\gamma$, relativistic beaming means that we can define the polar scattering angle $\phi'$ relative to the axis of the Lorentz boost.}
   \label{fig:ICS_LabRestFig}
\end{figure}

Inverse Compton scattering is the scattering of a photon to a higher energy off of an energetic electron.
As the photon can gain an appreciable fraction of the electron energy, the electron recoil effect must be taken into account.
It is convenient to treat ICS as Compton scattering in the rest-frame of the electron, as illustrated in fig.\,\ref {fig:ICS_LabRestFig}.
In terms of rest-frame quantities, the scattered photon energy $E_f'$ (for $a_0\ll1$) is given by,
\begin{align}
    \frac{E_f'}{E_i'} = \left({ 1+\frac{E_i'}{m_e c^2}\left(1-\cos\phi'\right)}\right)^{-1} \;,
    \label{eqn:restFrameScattered}
\end{align}
where $\phi'$ is the polar scattering angle and $E_i'$ is the incident photon energy.
The second term on the right-hand side of eq.\,\textup {(\ref {eqn:restFrameScattered})} represents the recoil of the electron, which significantly alters the resultant photon energy when $E_i'\sim m_e c^2$.

The incident photon energy in the electron rest-frame is found using the Lorentz transform of the lab-frame photon energy $E_i$, i.e.,
\begin{align}
    E_i' &= \gamma E_i\left(1-\beta\cos\theta\right) \;,
    \label{eqn:restFrameIncident}
\end{align}
where $c\beta$ is the electron velocity, $\theta$ is the angle between the electron and photon directions of motion in the lab-frame and $\gamma = 1/\sqrt{1-\beta^2}$.

With the assumption $\gamma\gg1$, relativistic beaming causes all incoming photons for which $\theta\gg1/\gamma$ to propagate anti-parallel to the original electron propagation direction when viewed in the rest-frame, i.e.~$\theta' \approx \pi$.
For the situations of interest in this paper, the condition $\theta\gg1/\gamma$ is always met and so we define $\phi'$ as relative to a constant incoming photon vector.
Using this definition, the scattered photon energy in the lab-frame is,
\begin{align}
    E_f &= \gamma E_f'(1-\beta\cos\phi') \;. \label{eqn:labFrameScattered}
\end{align}
For $\gamma \gg 1$, relativistic beaming causes all scattered photons to be emitted in approximately the electron direction of motion, i.e.~angles relative to the electron trajectory $\lesssim1/\gamma$.
Therefore, the scattered electron energy is calculated by subtracting the scattered photon energy from the initial electron energy.
Combining eqs.\,\textup {(\ref {eqn:restFrameScattered})},  \textup {(\ref {eqn:restFrameIncident})} and\, \textup {(\ref {eqn:labFrameScattered})} gives,
\begin{align}
    E_f &= \frac{\gamma^2 E_i\left(1-\beta\cos\theta\right)  (1-\beta\cos\phi')}{{ 1+{\gamma E_i\left(1-\beta\cos\theta\right)}\left(1-\cos\phi'\right)}/{m_e c^2}} \;.
    \label{eqn:ICS_photonEnergy}
\end{align}
The differential cross-section for Compton scattering for  $a_0\ll 1$ is given by the Klein-Nishina formula.
Averaging over the azimuthal angle $\psi'$, the differential cross-section is \cite{Berestetskii1982quantum},
\begin{align}
    \d{\sigma_{\it KN}}{\Omega} &= \frac{{r_e}^2}{2}\left(\frac{E'_f}{E'_i}\right)^2\left[\frac{E'_f}{E'_i}+\frac{E'_i}{E'_f} - \cos^2\phi'\right] \;,
    \label{eqn:KNformula}
\end{align}
where $r_e$ is the classical electron radius.
The total scattering cross-section  can be found by integrating eq.\,\textup {(\ref {eqn:KNformula})},
\begin{widetext}
\begin{align}
    \sigma_{\it KN} = \int_{0}^{2\pi}\int_{0}^{\pi} \d{\sigma_{\it KN}}{\Omega}\sin\phi'\mathrm{d}\phi'\mathrm{d}\psi' = \pi r_e^2 \left[\frac{(1+2x)^2-1}{2x(1+2x)^2} + \frac{\ln{(1+2x)}}{x}+\frac{4x-2(x+1)\ln(2x+1)}{x^3}\right] \;,
    \label{eqn:sigma_KN}
\end{align}
\end{widetext}
where $x=E_i'/m_e c^2$.

\begin{figure}[htb] 
   \centering
   \includegraphics[width=8.5cm]{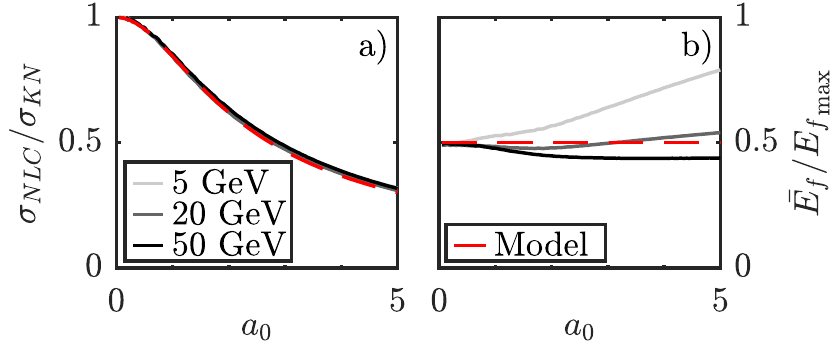} 
   \caption{Plot of a) total cross section and b) average photon energy calculated using \cite{Seipt2011PRA} for the given electron energies.
   The approximation used in the simplified model is plotted as the red dashed line for comparison.}
   \label{fig:ICS_SeiptModelComp}
\end{figure}

However, this simple picture, which is valid for low amplitudes, is complicated by non-linear effects for $a_0\gtrsim1$, such that the Klein-Nishina formulae are no longer accurate.
The electron motion in the ponderomotive potential of the laser pulse results in a time dependent redshift of the Compton scattered photons \cite{Seipt2015PRA} and modification of the differential cross-section.
In addition, harmonics of the fundamental scattering frequency are produced.
An approximate quantification of this effect was obtained by numerical fitting of the total scattering cross-section as calculated for circular polarisation using the framework established by Seipt and K\"ampfer \cite{Seipt2011PRA,Seipt2020PC}, as shown in fig.\,\ref {fig:ICS_SeiptModelComp}a.
For $0<a_0<5$, a good approximation to the non-linear Compton
scattering cross-section was found to be, 
\begin{align}
    \sigma_{\it NLC}(a_0) &\approx \frac{\sigma_{\it KN}}{\sqrt{1+0.4{a_0}^2}} & \text{(circular)} \;.
\end{align}
For linear polarisation, the cycle averaged value of $a{^2}$ should be used instead, i.e.,
\begin{align}
    \sigma_{\it NLC}(a_0) &\approx \frac{\sigma_{\it KN}}{\sqrt{1+0.2{a_0}^2}}  & \text{(linear)} \label{eqn:sigmaICS_lin} \;.
\end{align}
Although, the details of the scattered spectrum differ between linear and circular polarisation, this is not included for the approximate treatment used in this model.

The mean photon energy, averaged over all values of $\phi'$ approximates to half of the maximum possible photon energy, which occurs for $\phi'=\pi$, i.e.,
\begin{align}
    \bar{E_f} &\simeq  \frac{\gamma^2 E_i (1-\beta \cos\theta)} {1+\left[{2\gamma E_i(1-\beta \cos\theta)}/{m_e c^2}\right]} \;.
    \label{eqn:ICS_avg_energy}
\end{align}

For $a_0 \leq 5$ and \unit[$\gamma m_e ^2>2$]{GeV}, the average photon energy, calculated using \cite{Seipt2011PRA}, varies only slowly with $a_0$, varying by 20\% for $a_0=2$ compared to $a_0=0$, as shown in fig.\,\ref {fig:ICS_SeiptModelComp}b.
This is partly helped  by the contribution from higher harmonics balancing the ponderomotive reduction in the fundamental Compton energy.
Therefore, we apply the approximation of eq.\,\textup {(\ref {eqn:ICS_avg_energy})} independently of $a_0$, with the knowledge that the accuracy will be lower for $a_0>1$.

\section{Electron beam energy spread due to ICS} 

For a single electron, the expected number of scattering interactions, neglecting any change in cross-section due to the emission of multiple photons, from a linearly polarised laser with intensity distribution $I(t,\mathbf{x})$ is given by,
\begin{align}
f &=  \int_{t_0}^{t_1}\sigma_{\it NLC}\, \frac{I(t,\mathbf{x}(t))}{\hbar \omega_0}\, (1-\cos{\theta})\, \mathrm{d}t \;, \nonumber \\
f &\simeq \frac{\sigma_{\it KN}  \omega_0 m_e^2 c^3 \epsilon_0 \sin^2{\left(\frac{\theta}{2}\right)}}{\hbar e^2} \int_{t_0}^{t_1}  \frac{\hat{a}^2}{1+0.2\hat{a}^2} \, \mathrm{d}t \;,
\label{eqn:ICS_probability}
\end{align}
where $\hat{a}$ is the amplitude envelope of the normalised vector potential. 
The integral in eq.\,\textup {(\ref {eqn:ICS_probability})} is over the trajectory of the electron through the laser field from $\mathbf{x}(t_0)$ to  $\mathbf{x}(t_1)$.
With the approximation that all electrons have the same scattering probability, the fraction of electrons that undergo $n$ scattering interactions is given by the Poisson distribution with parameter $f$,
\begin{equation}
p_n =\frac{f^n e^{-f}}{n!} \;.
\label{eqn:pCalc}
\end{equation}

In order to determine the electron energy after $n$ interactions, we use the average emitted photon energy for ICS eq.\,\textup {(\ref {eqn:ICS_avg_energy})} and break up the emission of a photon into small pieces $\delta(\hbar \omega) = \delta n' \, \hbar \omega$, where $\delta n'$ is an infinitesimal part of the photon,  
\begin{align}
\delta (\gamma m_e c^2) &= -\frac{\gamma^2\,  \delta n' \, E_i (1-\beta \cos\theta)} {1+{(2\gamma \, \delta n' \, E_i (1-\beta \cos\theta)}/{m_e c^2})} \;.
\label{eqn:ICS_gammaDiff}
\end{align}
Taking the limit of $\delta n' \to 0$ and $\gamma \gg 1$ ($\beta\simeq1$), then eq.\,\textup {(\ref {eqn:ICS_gammaDiff})} can be integrated to find the electron energy after $n$ photon emissions,
\begin{align}
    \int_{\gamma_0}^{\gamma_n} \frac{ \mathrm{d}\gamma}{\gamma^2}&= -\frac{ 2 \sin^2\left(\frac{\theta}{2}\right) E_i}{m_e c^2} \int_0^n \mathrm{d}n' \;,
    \label{eqn:gamma_n_integral}
    \\
    \gamma_n &= \gamma_0\left(1+{n\gamma_0W}\right)^{-1} \;,
    \label{eqn:gamma_n}
\end{align}
where $W = 2 \sin^2\left(\frac{\theta}{2}\right) E_i/m_e c^2$.
It is interesting to note that although the integral in eq.\,\textup {(\ref {eqn:gamma_n_integral})} does not explicitly include the electron recoil, the integration yields the same values of $\gamma_n$ as those obtained by iteratively subtracting the photon energy given by eq.\,\textup {(\ref {eqn:ICS_avg_energy})}.

Using the above expressions, it is possible to analytically estimate the effect of ICS on the electron spectrum due to the photon recoil.
Electrons that lose too much energy may not be useful for subsequent application of the electron beam.
Therefore, one can calculate the fraction of electrons $F_a$ that remain within some acceptable energy spread $\Delta \gamma / \gamma_0$.
This can be estimated by summing the occupancies of each state for which the final electron energy remains within the acceptable energy range, i.e.,
\begin{equation}
F_a  = \sum_{n=0}^{n_{\mathrm{max}}} p_n \;,
\end{equation}
where,
\begin{equation}
    n_{\mathrm{max}} = \floor*{\frac{1}{\gamma_0 W} \left(\frac{{\Delta \gamma}/{\gamma_0}}{1-\Delta \gamma / \gamma_0}\right) }\;,
\end{equation}
where $\floor{ ~}$ is the floor operator.
For large electron energies, the energy lost from a single scattering interaction is already so large that a reasonable metric is the fraction of electrons that do not undergo scattering.
However, this neglects the possibility that electrons falling outside the desired energy may cause detrimental or damaging effects, in addition to reducing efficiency.
Instead, we use a measure of the energy spread as the normalised RMS deviation from the initial energy value, i.e., 
\begin{align}
\bar{\sigma}_{\gamma\mathrm{ICS}}&= \sqrt{ \sum_{n=0}^{\infty} p_n \left(\frac{\gamma_n-\gamma_0}{\gamma_0}\right)^2} \;. 
\label{eqn:relEnergySpread}
\end{align}
In practise, the sum in eq.\,\textup {(\ref {eqn:relEnergySpread})} can be limited to a value of $n$ at which the state occupation $p_n$ (using eq.\,\textup {(\ref {eqn:pCalc})}) becomes negligible.
For a beam with an initial energy spread $\bar{\sigma}_{\gamma0}$, the increase due to ICS is added in quadrature, i.e.~$\bar{\sigma}_\gamma = \sqrt{\bar{\sigma}_{\gamma0}^2 + \bar{\sigma}_{\gamma\mathrm{ICS}}^2}$.

A Monte-Carlo calculation was performed, in which \num{e5} electrons were initialised at an average energy \unit[50]{GeV} and 1\% RMS energy spread.
A \unit[$\lambda=1000$]{nm} Gaussian laser pulse with a duration of \unit[$\tau=112$]{fs} collided head on ($\theta=\pi$) and the scattering interactions were calculated in discrete time steps.
At each time step, the probability of each electron undergoing a scattering interaction was taken from a look-up table of total scattering cross-sections.
Electrons were randomly selected using these probabilities and the emitted photon energy was then similarly selected from a look-up table of the differential cross-sections.
Both look-up tables were created in advance using the method of Seipt and K\"ampfer \cite{Seipt2011PRA}.
The electron energy was then reduced by the emitted photon energy before proceeding to the next time step.

\begin{figure}[htb] 
   \centering
   \includegraphics[width=8.5cm]{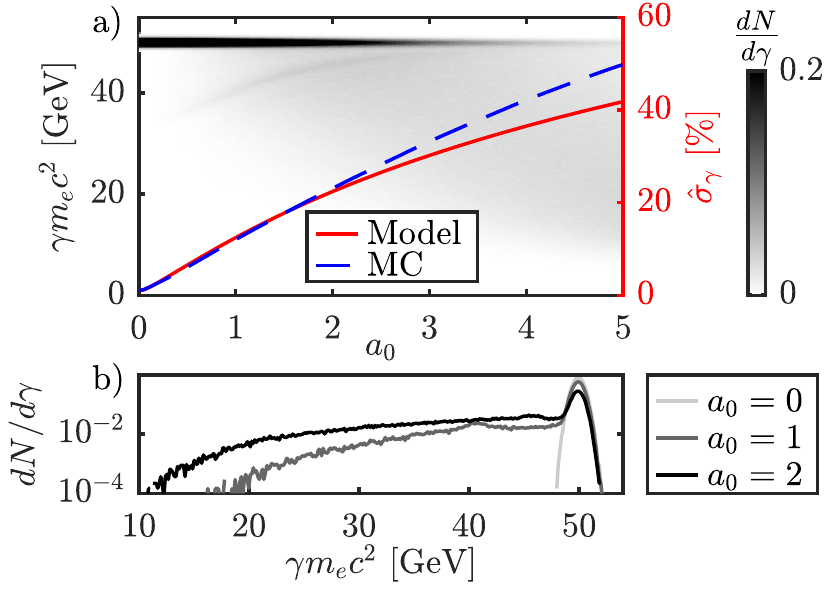} 
   \caption{a) Electron beam spectrum after \ang{90} collision with a Gaussian laser pulse from Monte-Carlo simulations with differing values of $a_0$.
   The initial electron beam had a mean energy of \unit[50]{GeV} with $\bar{\sigma}_{\gamma0} = 1\%$. 
   The resultant electron beam energy spread was calculated from the results of the Monte-Carlo simulation (blue) and the analytical model (red).
   b) Plots of the final electron spectrum for selected values of $a_0$.}
   \label{fig:ICS_ES_aScan}
\end{figure}

Figure\,\ref {fig:ICS_ES_aScan} shows the energy spectrum of the electron beam after collision with laser pulses with varying $a_0$.
As the average photon energy for the first emission is \unit[12]{GeV}, this is highly detrimental to the electron spectrum, even for low $a_0$. 
The energy spread of the electrons in the simulation is closely matched by the prediction of the analytical model, using eqs.\,\textup {(\ref {eqn:ICS_probability})},  \textup {(\ref {eqn:pCalc})},  \textup {(\ref {eqn:gamma_n})} and\, \textup {(\ref {eqn:relEnergySpread})}, for $a_0<2$.
The energy spread of the electron beam increases rapidly as a function of $a_0$, reaching $\approx50$\% for $a_0=5$.
For $a_0=1$, $14$\% of electrons undergo at least one scattering interaction.

\begin{figure}[ht] 
   \centering
   \includegraphics[width=8.5cm]{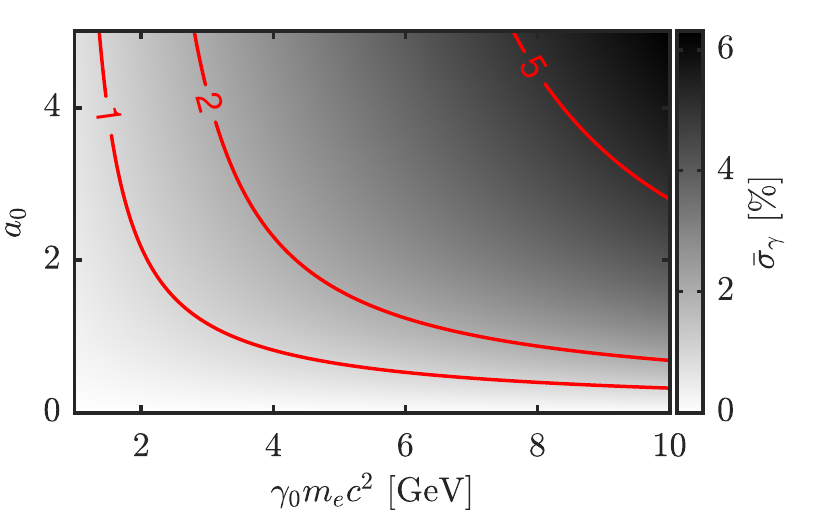} 
   \caption{Relative energy spread increase due to ICS at \ang{90} for varying $a_0$ and electron energy $\gamma m_e c^2$. 
   The Gaussian laser pulse had a wavelength of \unit[800]{nm}, \unit[$\tau = 50$]{fs} and infinite transverse extent. 
   The red isolines give the combination of electron energy and $a_0$ which results in the given percentage increase in energy spread..}
   \label{fig:ICS_energySpreadApprox_NL}
\end{figure}

At lower electron energies, photon recoil effects can still be important if a significant fraction of electrons undergo scattering reactions.
Figure\,\ref {fig:ICS_energySpreadApprox_NL} shows the predicted energy spread increase for an electron beam colliding with a Gaussian laser pulse at \ang{90}.
The alignment is such that the electron passes through the peak field of the laser, which has a pulse duration \unit[$\tau=50$]{fs}.
The energy spread increase is greatest for higher values of $a_0$ and $\gamma$.
This indicates that any collision with the extracted laser pulse at the exit of a multi-GeV LWFA must be at $a_0<1$ in order to achieve a final energy spread of $\bar{\sigma}_\gamma<0.01$.

The photons emitted by this process are determined by the electron and laser beam properties at the point of collision.
Therefore, diagnostics of the spectrum and spatial distribution of this photon source may be used to provide information about the electron beam and its collision with the laser at intermediate stages of a staged LWFA.

\section{Electron path through reflected laser field}

\begin{figure}[hpbt] 
   \centering
   \includegraphics[width=8.5cm]{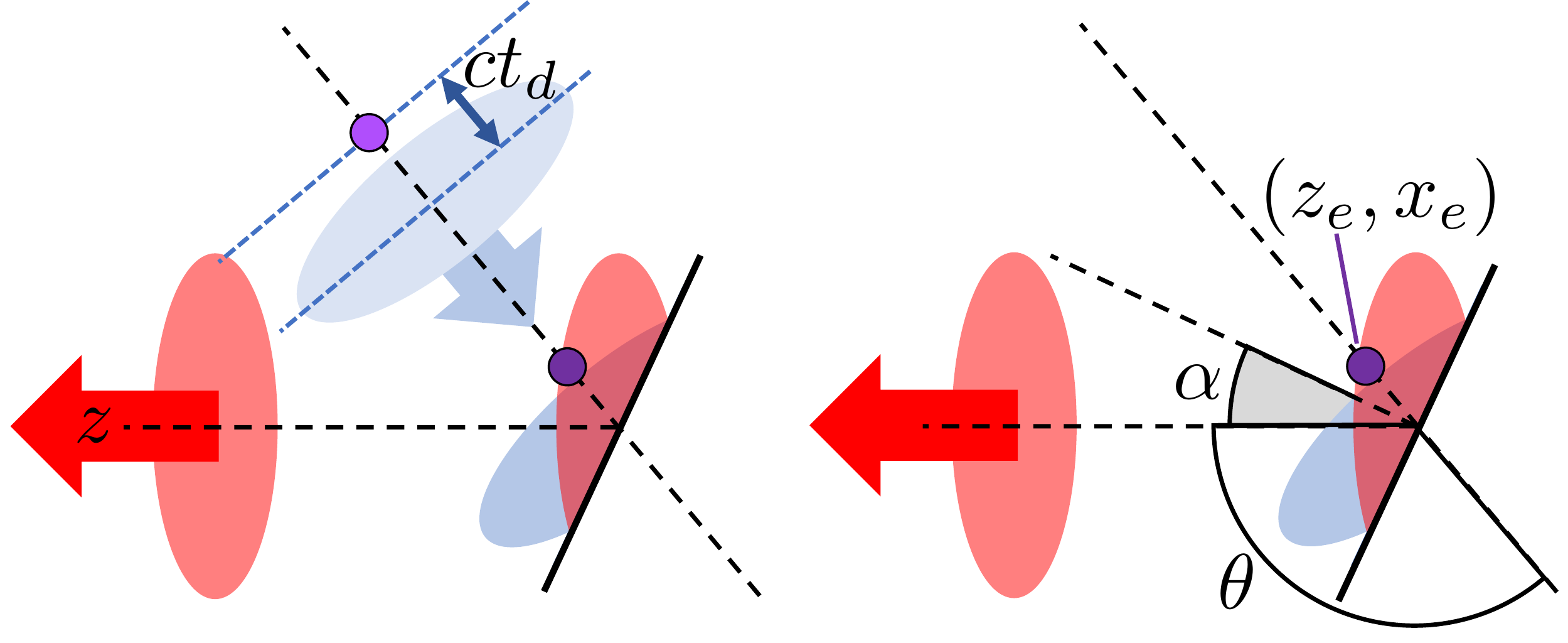} 
   \caption{Interaction geometry of electron beam (purple) trailing a laser pulse (blue). 
   The laser pulse after reflection (red) from the planar surface interacts with the electron bunch.}
   \label{fig:ICS_laserAOI}
\end{figure}

For the case of a laser reflecting from a plane (fig.\,\ref {fig:ICS_laserAOI}) with its surface normal at an angle $\alpha$ to the initial laser direction, then the electron collides with the reflected field at an angle $\theta=\pi-2\alpha$.

For simplicity, the reflected laser pulse envelope is defined as Gaussian in the longitudinal ($z$) and transverse ($x$) axes, 
\begin{align}
\hat{a}^2(t,x,z) &= {a_0}^2 \exp{\left[\frac{-2(z-ct)^2}{c^2 \tau^2}\right]}\exp{\left[\frac{-2x^2}{ {\sigma_x}^2}\right]} \;.
\label{eqn:aIntegral}
\end{align}
The electrons are assumed to occupy a single point, which moves collinearly with the centre of the laser pulse before reflection.
Their trajectory in the coordinate frame of the reflected laser is,
\begin{align}
    x_e &= c(t-t_d)\sin2\alpha \;, \nonumber \\
    z_e &= -c(t-t_d)\cos2\alpha  \;,
    \label{eqn:electronTrajectory}
\end{align}
where $t_d$ is the delay by which the electron bunch trailed the laser pulse before reflection.
Therefore, the electron experiences the field given by $\hat{a}(t,x_e,0)$ for $x_e<0$ and $a(t>t_d,x_e>0,0) =0$, i.e.~the field on the rear side of the plasma mirror is assumed to be zero.
Substituting eq.\,\textup {(\ref {eqn:electronTrajectory})} into eq.\,\textup {(\ref {eqn:aIntegral})} gives,
\begin{align}
\hat{a}^2(t,x_e,0) 
    &= {a_0}^2 A
    \exp\left[-2C_2\frac{(t-t_d C_1/C_2)^2}{\tau^2}\right] \;, 
    \label{eqn:electronAfield}
\end{align}
where,
\begin{align}
    A &=\exp\left[-\frac{2{t_d}^2}{\tau^2} \left(\cos^22\alpha  +\Phi^2\sin^22\alpha - \frac{{C_1}^2}{C_2}\right)\right] \;, \nonumber
    \\
    C_1 &= {\cos^22\alpha+ \cos2\alpha + \Phi^2\sin^22\alpha} \;, \nonumber \\
    C_2 &= {\cos^22\alpha +1 +2\cos2\alpha +\Phi^2\sin^22\alpha} \;.
\end{align}
and $\Phi =c \tau/\sigma_x$.
Equation\,\textup {(\ref {eqn:electronAfield})} is substituted into eq.\,\textup {(\ref {eqn:ICS_probability})}, with the limits $t_0 = -\infty$ and $t_0 = t_d$ in order to calculate $f$.
For collision with an orthogonally propagating laser pulse, without use of a plasma mirror, the upper limit on the integral in eq.\,\textup {(\ref {eqn:ICS_probability})} is replaced by $+\infty$.
The finite limit is due to the scattering process ending once the electron crosses the boundary of the foil at $t=t_d$.
The plasma mirror is assumed to be 100\% reflective in all of the following calculations.

\subsection{Self-guided highly non-linear regime}

\begin{figure}[hpbt] 
   \centering
   \includegraphics[width=8.5cm]{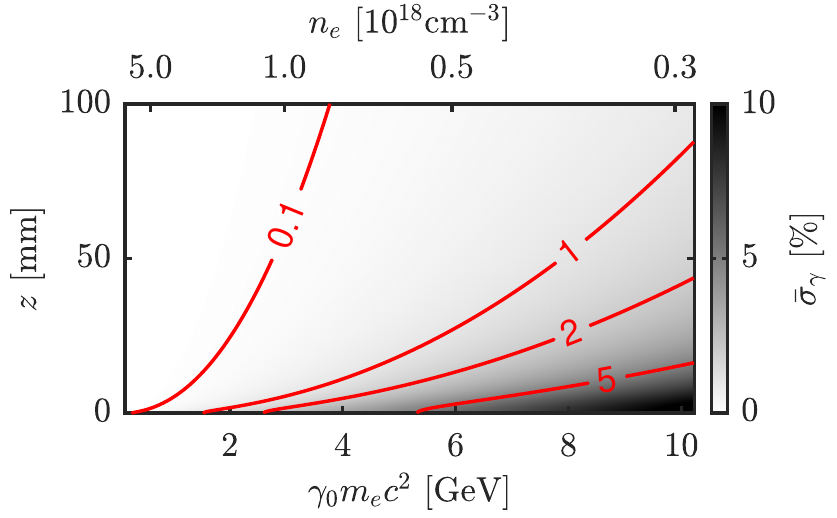} 
   \caption{Relative energy spread increase due to ICS with the drive laser of a `bubble' regime LWFA removed by a $45^{\circ}$ plasma mirror. The laser is allowed to freely diffract from the guided value of $a_0=5$ at $z=0$. 
   The required drift space to achieve a given percentage increase in energy spread as functions of electron energy are overlaid as red lines.}
   \label{fig:ICS_energySpreadFoil}
\end{figure}

For a self-guided, non-linear LWFA injector stage with electron density $n_e$, a matched laser pulse has $1/e^2$ radius given by $k_p\sigma_x \approx 2\sqrt{a_0}$ and the pulse duration $\omega_p\tau = \sqrt{a_0}$ \cite{Lu2007PRSTAB}.
As the laser exits the plasma it diffracts as $\sigma_x(z) = \sigma_x(0) \sqrt{1+(z/z_r)^2}$, where $z_r = \pi {\sigma_x}^2/\lambda$ for laser of wavelength $\lambda$.
The laser pulse amplitude decreases as the pulse diffracts as $a_0(z) =  a_0(0) /\sqrt{1+(z/z_r)^2}$.
At the point of dephasing, the electron beam trails the laser pulse by $t_d \sim \tau/2$.
To approximate the non-linear laser evolution in the plasma, we assume that pulse front etching has removed the front of the laser pulse up to the intensity peak.
This is included by setting the lower limit of integration $t_0 = t_d \cos{2\alpha}/(1+\cos{2\alpha})$.
The increased energy spread as functions of $z$ and $n_e$ was calculated by estimating $a_0(0) = 5$ and $\gamma_0 = (2/3)a_0 (n_c/n_e)$, where $n_c = \epsilon_0  m_e {\omega_0}^2 / e^2$.
The results are shown in fig.\,\ref {fig:ICS_energySpreadFoil}, showing the required drift length to achieve a specified relative energy spread for a given electron beam energy.

Due to the photon energy scaling in eq.\,\textup {(\ref {eqn:ICS_avg_energy})}, the electron recoil increases significantly at higher electron energies.
This can be mitigated by allowing the laser pulse to diffract and thereby reducing the peak intensity before extraction.
For \unit[$\gamma m_e c^2 = 5$]{GeV}, \unit[18]{mm} of free space after the plasma would be required in order to preserve a 1\% energy spread.
For higher electron energies or smaller target energy spread, the required drift space increases rapidly making the coupling between staged plasma accelerators longer and it may become impractical for beam quality to be preserved \cite{Floettmann2003}.

\subsection{Staged acceleration by quasi-linear LWFA}
\label{scn:StagedLWFA}

\begin{figure}[htb] 
   \centering
   \includegraphics[width=8.5cm]{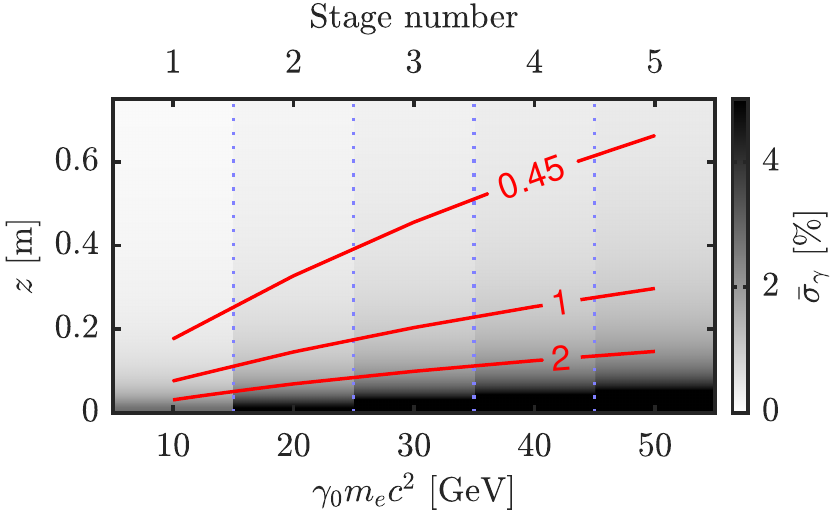} 
   \caption{Relative increase in energy spread per stage due to ICS with the drive laser removed by a $45^{\circ}$ plasma mirror.  
   The \unit[$\lambda=1$]{$\mu$m} laser is assumed to be externally guided in each stage with a matched spot size of \unit[$\sigma_x=70$]{$\mu$m} and $a_0=1.5$. Isolines of 1\% and 2\% relative energy spread increase per stage are marked alongside the required value (0.45\%) to achieve a final energy spread of 1\% at \unit[50]{GeV}.}
   \label{fig:ICS_energySpreadStaged}
\end{figure}

For a staged LWFA accelerator scheme, it may be advantageous to use externally guided quasi-linear LWFA stages,  with \unit[10]{GeV} energy gain per stage \cite{Schroeder2010PRSTAB}. 
The nominal laser properties are $a_0=1.5$, \unit[$\lambda=1$]{$\mu$m} and \unit[$\sigma_x = 70$]{$\mu$m} for \unit[$\sim1$]{m} long stages at \unit[$n_e = 10^{17}$]{cm$^{-3}$}.
To reach a final energy of \unit[50]{GeV}, five such stages are required, with the electron beam transported from the exit of each stage to the entrance of the next.
In order to obtain a final energy spread of $\hat{\sigma}_{\gamma}=1$\%, the energy spread increase at each stage must be limited to $<0.45$\%.
The relative energy spread increase at the end of each LWFA stage is given in fig.\,\ref {fig:ICS_energySpreadStaged} as a function of the drift distance $z$ between the plasma exit and the laser out-coupling.

In this case, a drift space of \unit[$z\approx 0.2$]{m} is required for the first stage rising to \unit[$z\approx 0.6$]{m} for the final stage where $\gamma=10^5$.
The large drift distances required to reduce the energy spread effect are a consequence of the large spot size at the plasma exit which results in a long Rayleigh range \unit[$z_R=15$]{mm}.
In reality, the requirements for a high luminosity collider may be much tighter than 1\% energy spread at the end of the accelerator, and so alternative concepts for laser beam extraction need to be considered.
At a distance of \unit[0.6]{m}, the laser spot width has diffracted to \unit[$\sigma_x=3$]{mm} with a peak intensity of \unit[$I_0\approx 2\times10^{15}$]{Wcm$^{-2}$}.
At this intensity, the plasma reflectively will be low \cite{Ziener2003JAP} and so a nominally reflective tape material is required.
Also, due to the transverse size of the beam, it will be possible to shape the reflecting surface so that off-axis laser radiation is not reflected into the path of the electron beam.
This, along with other mitigation strategies are discussed in the next section.

\section{Mitigation of energy spread growth due to ICS}

\subsection{Shaped reflector surface}
\begin{figure}[hpbt] 
   \centering
   \includegraphics[width=8.5cm]{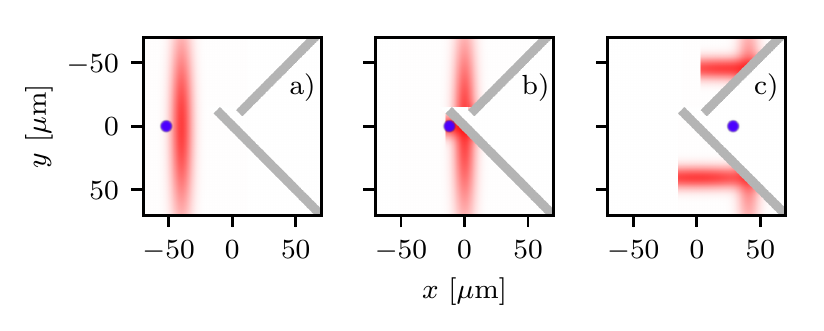} 
   \caption{Conceptual drawing for mitigating ICS energy spread increase through shaping of the reflecting surface. By deflecting as much energy as possible out of the path of the electron beam, it may be possible to reduce the collision overlap.}
   \label{fig:ICS_picMitigation}
\end{figure}

In order to mitigate the energy spread growth effect, it is most effective to try and minimise the interaction of the reflected laser field with the electron beam.
This can be achieved by shaping the reflecting surface so that the laser energy is entirely, or almost entirely, deflected away from the electron beam, as shown in fig.\,\ref {fig:ICS_picMitigation}.
This could be achieved by aligning the point of a triangular profile reflector, or the join of two flat foils, as close to the electron beam axis as possible.
Due to the finite size of the electron beam and alignment tolerances, it is not possible to completely eliminate the ICS interaction, but it can potentially be reduced.

\begin{figure}[htb] 
   \centering
   \includegraphics[width=8.5cm]{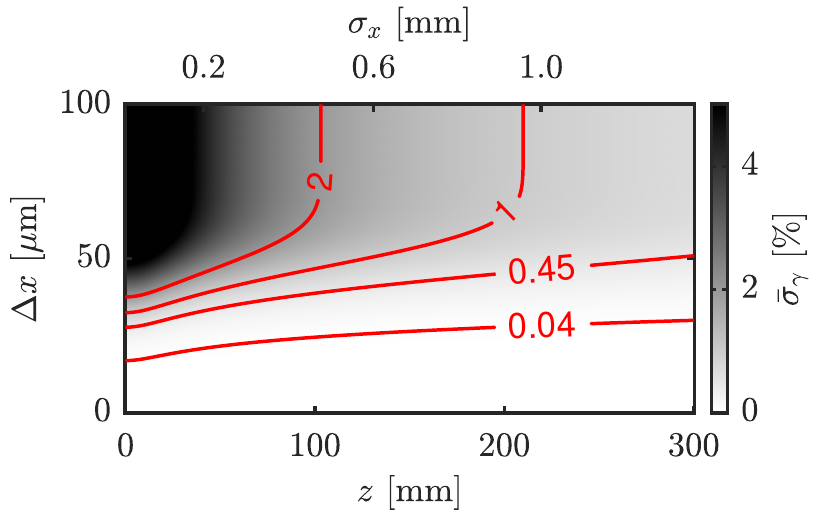} 
   \caption{Relative energy spread increase due to ICS at the output of the final stage of a staged \unit[50]{GeV} LWFA (see section \,\ref {scn:StagedLWFA}) using a shaped profile reflector. The corner of the profile is $\Delta{x}$ from the electron beam axis at a drift distance of $z$. The \unit[$\lambda=1$]{$\mu$m}, $a_0=1.5$ laser pulse had a transverse size \unit[$\sigma_{x}(0) = 70$]{$\mu$m} at the exit of a plasma of density \unit[$10^{17}$]{cm$^{-3}$}.}
   \label{fig:ICS_FoilLimited}
\end{figure}

The ICS energy spread effect can then be calculated by changing the lower limit of the integration in eq.\,\textup {(\ref {eqn:ICS_probability})} to $t_0=t_i$, when the electron beam first enters the laser field.
The value $t_i$ is determined by the precision to which the edge of the foil can be placed relative to the electron beam $\Delta{x}$, as $t_i = t_d - \Delta{x}/c$. 
Figure\,\ref {fig:ICS_FoilLimited} shows the reduction of the ICS energy spread effect depending on the alignment of the reflector corner to the electron beam axis.
In order to be of benefit, the foil must be aligned to within the spot width $\sigma_x$ and the pulse duration $c \tau$.
For drift distances of a few Rayleigh lengths, the required alignment precision is dominated by the pulse length, which for a matched laser pulse requires $\Delta{x} \lesssim \lambda_p/2$.
 
\subsection{Changing the angle of the reflector}
\begin{figure}[htb] 
   \centering
   \includegraphics[width=8.5cm]{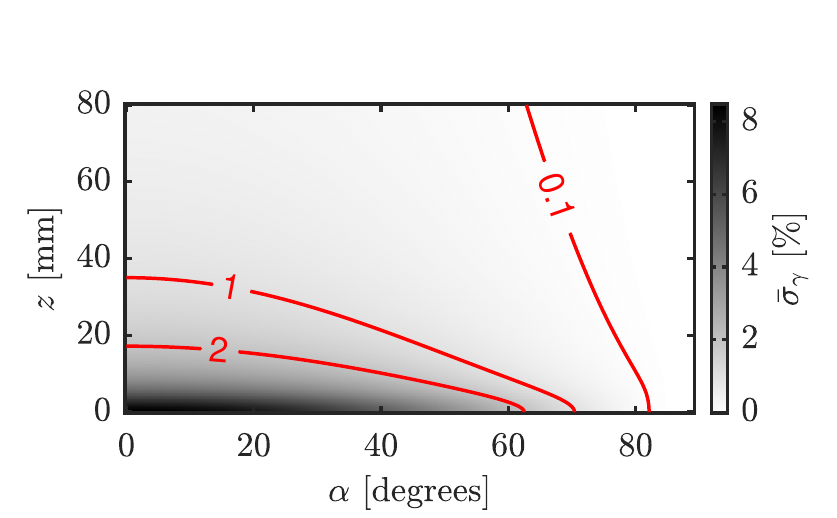}
   \caption{Energy spread increase as a function of reflector angle and drift distance from the exit of a \unit[5]{GeV} accelerator operating in the highly non-linear regime with $a_0=5$.}
   \label{fig:ICS_foilAngle}
\end{figure}

The fields experienced by the electron bunch and the energy of any photons produced can be altered by changing the angle at which the foil reflects the laser.
In particular, using a shallower angle of incidence reduces the generated photon energy due to the angular dependence in eq.\,\textup {(\ref {eqn:restFrameIncident})}.
The model was used to calculate the energy spread increase at the exit of a highly non-linear \unit[5]{GeV} LWFA stage, with the same parameters as fig.\,\ref {fig:ICS_energySpreadFoil}.
The results of the calculation, shown in fig.\,\ref {fig:ICS_foilAngle}, show that the ICS energy spread increase is indeed reduced for shallower angles.
In these calculations we have not included the additional divergence that occurs during the interaction, which would further reduce the scattering probability for large values of $\alpha$.
Therefore, increasing the reflector angle may be used to mitigate this effect provided it is allowed by the coupling geometry.
However, this comes at a cost of increasing the apparent thickness of the reflector material to the electron beam, which was not included in this analysis.

\subsection{Electron acceleration in later plasma periods}

\begin{figure}[htb] 
   \centering
   \includegraphics[width=8.5cm]{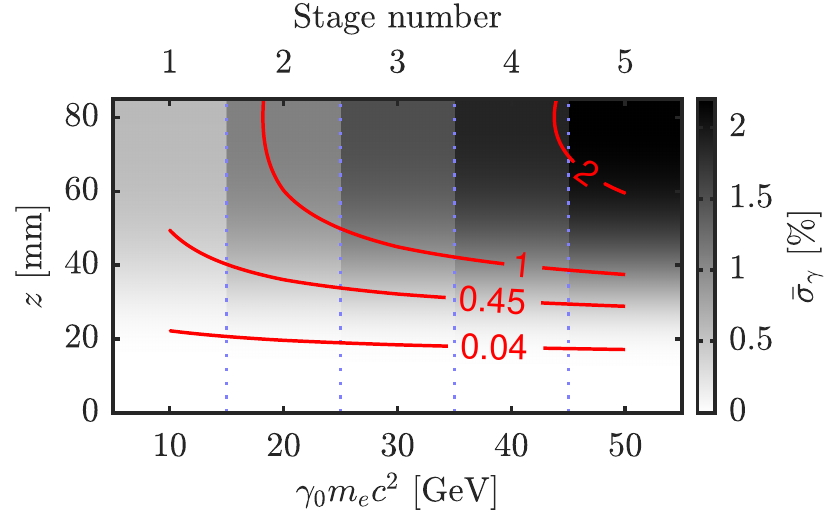}
   \caption{Relative increase in energy spread per stage due to ICS with the drive laser when accelerating in the 3\textsuperscript{rd} plasma period.  
   The \unit[$\lambda=1$]{$\mu$m} laser is assumed to be externally guided in each stage with \unit[$\sigma_x=70$]{$\mu$m} and $a_0=1.5$. Contours of 1\% and 2\% relative energy spread increase per stage are marked alongside the required values (0.45\% and 0.04\%) to achieve a final energy spread of 1\% and 0.1\% at \unit[50]{GeV}.}
   \label{fig:ICS_staged2ndBucket}
\end{figure}

A particularly effective mitigation strategy for the quasi-linear regime, i.e.~for \unit[10]{GeV} acceleration stages \cite{Schroeder2010PRSTAB}, would be to accelerate the electron beam in the second (or further) period of the plasma wave.
This increases the temporal separation between the laser and the electron bunch and such that a collision can be entirely avoided, as long as the transverse size of the laser is small, i.e.~$\sigma_x<(m+0.5) \lambda_p$ for the $m^{\mathrm{th}}$ plasma period.
In this case, a small drift distance is required, so that the laser is removed before it can significantly diffract.
This is shown in fig.\,\ref {fig:ICS_staged2ndBucket}, again for the case of a staged \unit[50]{GeV} accelerator with acceleration in the third plasma period.
For this method, the drift distance must be \unit[$<40$]{mm} in order to maintain a sub 1\% relative energy spread after the final stage.

Acceleration of electrons after one or more plasma oscillation periods may have other drawbacks, particularly as it allows for more non-linear development of the plasma wave. 
However, as it prevents any interaction between the drive laser and the electron beam, either by ICS during laser extraction or by direct laser acceleration during the acceleration itself \cite{Mangles2006PRL}, there may be significant advantages in terms of beam quality.

\subsection{Alternative staged LWFA concepts}
The multiple pulse LWFA scheme \cite{Hooker2014JPB}, proposes to use multiple laser pulses, spaced at intervals of the plasma period in order to resonantly drive the plasma wave.
Each individual pulse has a lower intensity $a_0\ll1$, but they act together to increase the amplitude of the plasma wave with each successive period.
An electron beam is then injected into the wake after the train of laser pulses.
In this scheme, the electron beam is by design placed in a plasma period away from all of the laser energy.
Also, each individual laser pulse is at a lower intensity and so any collision will have less effect.
Thus, using this scheme would avoid any issues with collision between the laser fields and the accelerated electron bunch.

One can also consider alternatives to plasma-mirror based LWFA staging. On example is to use curved plasma channels both for in-coupling and out-coupling the driving laser pulses from a straight plasma accelerator \cite{Luo2018PRL}.
In this case, it may be possible to avoid direct laser-electron beam interaction at the staging points and at the eventual plasma exit.

\section{Conclusion}
In this paper, we have investigated the potential increase in the energy spread of an electron beam due to inverse Compton scattering in laser-driven plasma wakefield accelerators.
The problem has shown to be particularly severe for high final electron energies, as the electrons can emit photons with a significant fraction of their energy when oscillating in the field of the extracted laser pulse.
Laser pulse extraction is a necessity for staged accelerators, and so this effect will be a serious factor for applications that require low energy spread such as \unit[$\sim5$]{GeV} FEL \cite{Walker2017JP} and \unit[$\sim50$]{GeV} high energy physics \cite{Wing2019RS} facilities.
Sub-percent energy spread beams from these devices will only be possible with careful extraction of the driving laser pulse, or by accelerating in the second or third period in the case of quasi-linear accelerator stages.

\section{Acknowledgements}
MS thanks D.\,Seipt, E.\,Gerstmayr, R.\,Watt and B.\,Appelbe for valuable discussions and D.\,Seipt for calculations.
We acknowledge funding from Science and Technology Facilities Council
Grant No. ST/P002021/1 and the EU Horizon 2020 research and innovation programme grant No.~653782.


\begin{thebibliography}{34}%
\makeatletter
\providecommand \@ifxundefined [1]{%
 \@ifx{#1\undefined}
}%
\providecommand \@ifnum [1]{%
 \ifnum #1\expandafter \@firstoftwo
 \else \expandafter \@secondoftwo
 \fi
}%
\providecommand \@ifx [1]{%
 \ifx #1\expandafter \@firstoftwo
 \else \expandafter \@secondoftwo
 \fi
}%
\providecommand \natexlab [1]{#1}%
\providecommand \enquote  [1]{``#1''}%
\providecommand \bibnamefont  [1]{#1}%
\providecommand \bibfnamefont [1]{#1}%
\providecommand \citenamefont [1]{#1}%
\providecommand \href@noop [0]{\@secondoftwo}%
\providecommand \href [0]{\begingroup \@sanitize@url \@href}%
\providecommand \@href[1]{\@@startlink{#1}\@@href}%
\providecommand \@@href[1]{\endgroup#1\@@endlink}%
\providecommand \@sanitize@url [0]{\catcode `\\12\catcode `\$12\catcode
  `\&12\catcode `\#12\catcode `\^12\catcode `\_12\catcode `\%12\relax}%
\providecommand \@@startlink[1]{}%
\providecommand \@@endlink[0]{}%
\providecommand \url  [0]{\begingroup\@sanitize@url \@url }%
\providecommand \@url [1]{\endgroup\@href {#1}{\urlprefix }}%
\providecommand \urlprefix  [0]{URL }%
\providecommand \Eprint [0]{\href }%
\providecommand \doibase [0]{http://dx.doi.org/}%
\providecommand \selectlanguage [0]{\@gobble}%
\providecommand \bibinfo  [0]{\@secondoftwo}%
\providecommand \bibfield  [0]{\@secondoftwo}%
\providecommand \translation [1]{[#1]}%
\providecommand \BibitemOpen [0]{}%
\providecommand \bibitemStop [0]{}%
\providecommand \bibitemNoStop [0]{.\EOS\space}%
\providecommand \EOS [0]{\spacefactor3000\relax}%
\providecommand \BibitemShut  [1]{\csname bibitem#1\endcsname}%
\let\auto@bib@innerbib\@empty
\bibitem [{\citenamefont {Tajima}\ and\ \citenamefont
  {Dawson}(1979)}]{Tajima1979PRL}%
  \BibitemOpen
  \bibfield  {author} {\bibinfo {author} {\bibfnamefont {T.}~\bibnamefont
  {Tajima}}\ and\ \bibinfo {author} {\bibfnamefont {J.~M.}\ \bibnamefont
  {Dawson}},\ }\href {http://link.aps.org/doi/10.1103/PhysRevLett.43.267}
  {\bibfield  {journal} {\bibinfo  {journal} {Physical Review Letters}\
  }\textbf {\bibinfo {volume} {43}},\ \bibinfo {pages} {267} (\bibinfo {year}
  {1979})}\BibitemShut {NoStop}%
\bibitem [{\citenamefont {Mangles}\ \emph {et~al.}(2004)\citenamefont
  {Mangles}, \citenamefont {Murphy}, \citenamefont {Najmudin}, \citenamefont
  {Thomas}, \citenamefont {Collier}, \citenamefont {Dangor}, \citenamefont
  {Divall}, \citenamefont {Foster}, \citenamefont {Gallacher}, \citenamefont
  {Hooker}, \citenamefont {Jaroszynski}, \citenamefont {Langley}, \citenamefont
  {Mori}, \citenamefont {Norreys}, \citenamefont {Tsung}, \citenamefont
  {Viskup}, \citenamefont {Walton},\ and\ \citenamefont
  {Krushelnick}}]{Mangles2004N}%
  \BibitemOpen
  \bibfield  {author} {\bibinfo {author} {\bibfnamefont {S.~P.~D.}\
  \bibnamefont {Mangles}}, \bibinfo {author} {\bibfnamefont {C.~D.}\
  \bibnamefont {Murphy}}, \bibinfo {author} {\bibfnamefont {Z.}~\bibnamefont
  {Najmudin}}, \bibinfo {author} {\bibfnamefont {A.~G.~R.}\ \bibnamefont
  {Thomas}}, \bibinfo {author} {\bibfnamefont {J.~L.}\ \bibnamefont {Collier}},
  \bibinfo {author} {\bibfnamefont {A.~E.}\ \bibnamefont {Dangor}}, \bibinfo
  {author} {\bibfnamefont {E.~J.}\ \bibnamefont {Divall}}, \bibinfo {author}
  {\bibfnamefont {P.~S.}\ \bibnamefont {Foster}}, \bibinfo {author}
  {\bibfnamefont {J.~G.}\ \bibnamefont {Gallacher}}, \bibinfo {author}
  {\bibfnamefont {C.~J.}\ \bibnamefont {Hooker}}, \bibinfo {author}
  {\bibfnamefont {D.~A.}\ \bibnamefont {Jaroszynski}}, \bibinfo {author}
  {\bibfnamefont {A.~J.}\ \bibnamefont {Langley}}, \bibinfo {author}
  {\bibfnamefont {W.~B.}\ \bibnamefont {Mori}}, \bibinfo {author}
  {\bibfnamefont {P.~A.}\ \bibnamefont {Norreys}}, \bibinfo {author}
  {\bibfnamefont {F.~S.}\ \bibnamefont {Tsung}}, \bibinfo {author}
  {\bibfnamefont {R.}~\bibnamefont {Viskup}}, \bibinfo {author} {\bibfnamefont
  {B.~R.}\ \bibnamefont {Walton}}, \ and\ \bibinfo {author} {\bibfnamefont
  {K.}~\bibnamefont {Krushelnick}},\ }\href {\doibase 10.1038/nature02939}
  {\bibfield  {journal} {\bibinfo  {journal} {Nature}\ }\textbf {\bibinfo
  {volume} {431}},\ \bibinfo {pages} {535} (\bibinfo {year}
  {2004})}\BibitemShut {NoStop}%
\bibitem [{\citenamefont {Faure}\ \emph {et~al.}(2004)\citenamefont {Faure},
  \citenamefont {Glinec}, \citenamefont {Pukhov}, \citenamefont {Kiselev},
  \citenamefont {Gordienko}, \citenamefont {Lefebvre}, \citenamefont
  {Rousseau}, \citenamefont {Burgy},\ and\ \citenamefont {Malka}}]{Faure2004N}%
  \BibitemOpen
  \bibfield  {author} {\bibinfo {author} {\bibfnamefont {J.}~\bibnamefont
  {Faure}}, \bibinfo {author} {\bibfnamefont {Y.}~\bibnamefont {Glinec}},
  \bibinfo {author} {\bibfnamefont {A.}~\bibnamefont {Pukhov}}, \bibinfo
  {author} {\bibfnamefont {S.}~\bibnamefont {Kiselev}}, \bibinfo {author}
  {\bibfnamefont {S.}~\bibnamefont {Gordienko}}, \bibinfo {author}
  {\bibfnamefont {E.}~\bibnamefont {Lefebvre}}, \bibinfo {author}
  {\bibfnamefont {J.-P.}\ \bibnamefont {Rousseau}}, \bibinfo {author}
  {\bibfnamefont {F.}~\bibnamefont {Burgy}}, \ and\ \bibinfo {author}
  {\bibfnamefont {V.}~\bibnamefont {Malka}},\ }\href {\doibase
  10.1038/nature02963} {\bibfield  {journal} {\bibinfo  {journal} {Nature}\
  }\textbf {\bibinfo {volume} {431}},\ \bibinfo {pages} {541} (\bibinfo {year}
  {2004})}\BibitemShut {NoStop}%
\bibitem [{\citenamefont {Geddes}\ \emph {et~al.}(2004)\citenamefont {Geddes},
  \citenamefont {Toth}, \citenamefont {Van~Tilborg}, \citenamefont {Esarey},
  \citenamefont {Schroeder}, \citenamefont {Bruhwiler}, \citenamefont {Nieter},
  \citenamefont {Cary},\ and\ \citenamefont {Leemans}}]{Geddes2004N}%
  \BibitemOpen
  \bibfield  {author} {\bibinfo {author} {\bibfnamefont {C.~G.~R.}\
  \bibnamefont {Geddes}}, \bibinfo {author} {\bibfnamefont {C.~S.}\
  \bibnamefont {Toth}}, \bibinfo {author} {\bibfnamefont {J.}~\bibnamefont
  {Van~Tilborg}}, \bibinfo {author} {\bibfnamefont {E.}~\bibnamefont {Esarey}},
  \bibinfo {author} {\bibfnamefont {C.~B.}\ \bibnamefont {Schroeder}}, \bibinfo
  {author} {\bibfnamefont {D.}~\bibnamefont {Bruhwiler}}, \bibinfo {author}
  {\bibfnamefont {C.}~\bibnamefont {Nieter}}, \bibinfo {author} {\bibfnamefont
  {J.}~\bibnamefont {Cary}}, \ and\ \bibinfo {author} {\bibfnamefont {W.~P.}\
  \bibnamefont {Leemans}},\ }\href {\doibase 10.1038/nature02900} {\bibfield
  {journal} {\bibinfo  {journal} {Nature}\ }\textbf {\bibinfo {volume} {431}},\
  \bibinfo {pages} {538} (\bibinfo {year} {2004})}\BibitemShut {NoStop}%
\bibitem [{\citenamefont {Leemans}\ and\ \citenamefont
  {Esarey}(2009)}]{Leemans2009PT}%
  \BibitemOpen
  \bibfield  {author} {\bibinfo {author} {\bibfnamefont {W.}~\bibnamefont
  {Leemans}}\ and\ \bibinfo {author} {\bibfnamefont {E.}~\bibnamefont
  {Esarey}},\ }\href {\doibase 10.1063/1.3099645} {\bibfield  {journal}
  {\bibinfo  {journal} {Physics Today}\ }\textbf {\bibinfo {volume} {62}},\
  \bibinfo {pages} {44} (\bibinfo {year} {2009})}\BibitemShut {NoStop}%
\bibitem [{\citenamefont {Steinke}\ \emph {et~al.}(2016)\citenamefont
  {Steinke}, \citenamefont {Van~Tilborg}, \citenamefont {Benedetti},
  \citenamefont {Geddes}, \citenamefont {Schroeder}, \citenamefont {Daniels},
  \citenamefont {Swanson}, \citenamefont {Gonsalves}, \citenamefont {Nakamura},
  \citenamefont {Matlis}, \citenamefont {Shaw}, \citenamefont {Esarey},\ and\
  \citenamefont {Leemans}}]{Steinke2016N}%
  \BibitemOpen
  \bibfield  {author} {\bibinfo {author} {\bibfnamefont {S.}~\bibnamefont
  {Steinke}}, \bibinfo {author} {\bibfnamefont {J.}~\bibnamefont
  {Van~Tilborg}}, \bibinfo {author} {\bibfnamefont {C.}~\bibnamefont
  {Benedetti}}, \bibinfo {author} {\bibfnamefont {C.~G.}\ \bibnamefont
  {Geddes}}, \bibinfo {author} {\bibfnamefont {C.~B.}\ \bibnamefont
  {Schroeder}}, \bibinfo {author} {\bibfnamefont {J.}~\bibnamefont {Daniels}},
  \bibinfo {author} {\bibfnamefont {K.~K.}\ \bibnamefont {Swanson}}, \bibinfo
  {author} {\bibfnamefont {A.~J.}\ \bibnamefont {Gonsalves}}, \bibinfo {author}
  {\bibfnamefont {K.}~\bibnamefont {Nakamura}}, \bibinfo {author}
  {\bibfnamefont {N.~H.}\ \bibnamefont {Matlis}}, \bibinfo {author}
  {\bibfnamefont {B.~H.}\ \bibnamefont {Shaw}}, \bibinfo {author}
  {\bibfnamefont {E.}~\bibnamefont {Esarey}}, \ and\ \bibinfo {author}
  {\bibfnamefont {W.~P.}\ \bibnamefont {Leemans}},\ }\href {\doibase
  10.1038/nature16525} {\bibfield  {journal} {\bibinfo  {journal} {Nature}\
  }\textbf {\bibinfo {volume} {530}},\ \bibinfo {pages} {190} (\bibinfo {year}
  {2016})}\BibitemShut {NoStop}%
\bibitem [{\citenamefont {Wing}(2019)}]{Wing2019RS}%
  \BibitemOpen
  \bibfield  {author} {\bibinfo {author} {\bibfnamefont {M.}~\bibnamefont
  {Wing}},\ }\href {\doibase 10.1098/rsta.2018.0185} {\bibfield  {journal}
  {\bibinfo  {journal} {Philosophical Transactions of the Royal Society A:
  Mathematical, Physical and Engineering Sciences}\ }\textbf {\bibinfo {volume}
  {377}},\ \bibinfo {pages} {20180185} (\bibinfo {year} {2019})}\BibitemShut
  {NoStop}%
\bibitem [{\citenamefont {Streeter}\ \emph {et~al.}(2018)\citenamefont
  {Streeter}, \citenamefont {Kneip}, \citenamefont {Bloom}, \citenamefont
  {Bendoyro}, \citenamefont {Chekhlov}, \citenamefont {Dangor}, \citenamefont
  {D{\"{o}}pp}, \citenamefont {Hooker}, \citenamefont {Holloway}, \citenamefont
  {Jiang}, \citenamefont {Lopes}, \citenamefont {Nakamura}, \citenamefont
  {Norreys}, \citenamefont {Palmer}, \citenamefont {Rajeev}, \citenamefont
  {Schreiber}, \citenamefont {Symes}, \citenamefont {Wing}, \citenamefont
  {Mangles},\ and\ \citenamefont {Najmudin}}]{Streeter2018PRL}%
  \BibitemOpen
  \bibfield  {author} {\bibinfo {author} {\bibfnamefont {M.~J.~V.}\
  \bibnamefont {Streeter}}, \bibinfo {author} {\bibfnamefont {S.}~\bibnamefont
  {Kneip}}, \bibinfo {author} {\bibfnamefont {M.~S.}\ \bibnamefont {Bloom}},
  \bibinfo {author} {\bibfnamefont {R.~A.}\ \bibnamefont {Bendoyro}}, \bibinfo
  {author} {\bibfnamefont {O.}~\bibnamefont {Chekhlov}}, \bibinfo {author}
  {\bibfnamefont {A.~E.}\ \bibnamefont {Dangor}}, \bibinfo {author}
  {\bibfnamefont {A.}~\bibnamefont {D{\"{o}}pp}}, \bibinfo {author}
  {\bibfnamefont {C.~J.}\ \bibnamefont {Hooker}}, \bibinfo {author}
  {\bibfnamefont {J.}~\bibnamefont {Holloway}}, \bibinfo {author}
  {\bibfnamefont {J.}~\bibnamefont {Jiang}}, \bibinfo {author} {\bibfnamefont
  {N.~C.}\ \bibnamefont {Lopes}}, \bibinfo {author} {\bibfnamefont
  {H.}~\bibnamefont {Nakamura}}, \bibinfo {author} {\bibfnamefont {P.~A.}\
  \bibnamefont {Norreys}}, \bibinfo {author} {\bibfnamefont {C.~A.~J.}\
  \bibnamefont {Palmer}}, \bibinfo {author} {\bibfnamefont {P.~P.}\
  \bibnamefont {Rajeev}}, \bibinfo {author} {\bibfnamefont {J.}~\bibnamefont
  {Schreiber}}, \bibinfo {author} {\bibfnamefont {D.~R.}\ \bibnamefont
  {Symes}}, \bibinfo {author} {\bibfnamefont {M.}~\bibnamefont {Wing}},
  \bibinfo {author} {\bibfnamefont {S.~P.~D.}\ \bibnamefont {Mangles}}, \ and\
  \bibinfo {author} {\bibfnamefont {Z.}~\bibnamefont {Najmudin}},\ }\href
  {\doibase 10.1103/PhysRevLett.120.254801} {\bibfield  {journal} {\bibinfo
  {journal} {Physical Review Letters}\ }\textbf {\bibinfo {volume} {120}},\
  \bibinfo {pages} {254801} (\bibinfo {year} {2018})}\BibitemShut {NoStop}%
\bibitem [{\citenamefont {Ziener}\ \emph {et~al.}(2003)\citenamefont {Ziener},
  \citenamefont {Foster}, \citenamefont {Divall}, \citenamefont {Hooker},
  \citenamefont {Hutchinson}, \citenamefont {Langley},\ and\ \citenamefont
  {Neely}}]{Ziener2003JAP}%
  \BibitemOpen
  \bibfield  {author} {\bibinfo {author} {\bibfnamefont {C.}~\bibnamefont
  {Ziener}}, \bibinfo {author} {\bibfnamefont {P.~S.}\ \bibnamefont {Foster}},
  \bibinfo {author} {\bibfnamefont {E.~J.}\ \bibnamefont {Divall}}, \bibinfo
  {author} {\bibfnamefont {C.~J.}\ \bibnamefont {Hooker}}, \bibinfo {author}
  {\bibfnamefont {M.~H.~R.}\ \bibnamefont {Hutchinson}}, \bibinfo {author}
  {\bibfnamefont {A.~J.}\ \bibnamefont {Langley}}, \ and\ \bibinfo {author}
  {\bibfnamefont {D.}~\bibnamefont {Neely}},\ }\href {\doibase
  10.1063/1.1525062} {\bibfield  {journal} {\bibinfo  {journal} {Journal of
  Applied Physics}\ }\textbf {\bibinfo {volume} {93}},\ \bibinfo {pages} {768}
  (\bibinfo {year} {2003})}\BibitemShut {NoStop}%
\bibitem [{\citenamefont {Ta~Phuoc}\ \emph {et~al.}(2012)\citenamefont
  {Ta~Phuoc}, \citenamefont {Corde}, \citenamefont {Thaury}, \citenamefont
  {Malka}, \citenamefont {Tafzi}, \citenamefont {Goddet}, \citenamefont {Shah},
  \citenamefont {Sebban},\ and\ \citenamefont {Rousse}}]{TaPhuoc2012NP}%
  \BibitemOpen
  \bibfield  {author} {\bibinfo {author} {\bibfnamefont {K.}~\bibnamefont
  {Ta~Phuoc}}, \bibinfo {author} {\bibfnamefont {S.}~\bibnamefont {Corde}},
  \bibinfo {author} {\bibfnamefont {C.}~\bibnamefont {Thaury}}, \bibinfo
  {author} {\bibfnamefont {V.}~\bibnamefont {Malka}}, \bibinfo {author}
  {\bibfnamefont {A.}~\bibnamefont {Tafzi}}, \bibinfo {author} {\bibfnamefont
  {J.~P.}\ \bibnamefont {Goddet}}, \bibinfo {author} {\bibfnamefont {R.~C.}\
  \bibnamefont {Shah}}, \bibinfo {author} {\bibfnamefont {S.}~\bibnamefont
  {Sebban}}, \ and\ \bibinfo {author} {\bibfnamefont {A.}~\bibnamefont
  {Rousse}},\ }\href {\doibase 10.1038/nphoton.2012.82} {\bibfield  {journal}
  {\bibinfo  {journal} {Nature Photonics}\ }\textbf {\bibinfo {volume} {6}},\
  \bibinfo {pages} {308} (\bibinfo {year} {2012})}\BibitemShut {NoStop}%
\bibitem [{\citenamefont {Cole}\ \emph {et~al.}(2018)\citenamefont {Cole},
  \citenamefont {Behm}, \citenamefont {Gerstmayr}, \citenamefont {Blackburn},
  \citenamefont {Wood}, \citenamefont {Baird}, \citenamefont {Duff},
  \citenamefont {Harvey}, \citenamefont {Ilderton}, \citenamefont {Joglekar},
  \citenamefont {Krushelnick}, \citenamefont {Kuschel}, \citenamefont
  {Marklund}, \citenamefont {McKenna}, \citenamefont {Murphy}, \citenamefont
  {Poder}, \citenamefont {Ridgers}, \citenamefont {Samarin}, \citenamefont
  {Sarri}, \citenamefont {Symes}, \citenamefont {Thomas}, \citenamefont
  {Warwick}, \citenamefont {Zepf}, \citenamefont {Najmudin},\ and\
  \citenamefont {Mangles}}]{Cole2018PRX}%
  \BibitemOpen
  \bibfield  {author} {\bibinfo {author} {\bibfnamefont {J.~M.}\ \bibnamefont
  {Cole}}, \bibinfo {author} {\bibfnamefont {K.~T.}\ \bibnamefont {Behm}},
  \bibinfo {author} {\bibfnamefont {E.}~\bibnamefont {Gerstmayr}}, \bibinfo
  {author} {\bibfnamefont {T.~G.}\ \bibnamefont {Blackburn}}, \bibinfo {author}
  {\bibfnamefont {J.~C.}\ \bibnamefont {Wood}}, \bibinfo {author}
  {\bibfnamefont {C.~D.}\ \bibnamefont {Baird}}, \bibinfo {author}
  {\bibfnamefont {M.~J.}\ \bibnamefont {Duff}}, \bibinfo {author}
  {\bibfnamefont {C.}~\bibnamefont {Harvey}}, \bibinfo {author} {\bibfnamefont
  {A.}~\bibnamefont {Ilderton}}, \bibinfo {author} {\bibfnamefont {A.~S.}\
  \bibnamefont {Joglekar}}, \bibinfo {author} {\bibfnamefont {K.}~\bibnamefont
  {Krushelnick}}, \bibinfo {author} {\bibfnamefont {S.}~\bibnamefont
  {Kuschel}}, \bibinfo {author} {\bibfnamefont {M.}~\bibnamefont {Marklund}},
  \bibinfo {author} {\bibfnamefont {P.}~\bibnamefont {McKenna}}, \bibinfo
  {author} {\bibfnamefont {C.~D.}\ \bibnamefont {Murphy}}, \bibinfo {author}
  {\bibfnamefont {K.}~\bibnamefont {Poder}}, \bibinfo {author} {\bibfnamefont
  {C.~P.}\ \bibnamefont {Ridgers}}, \bibinfo {author} {\bibfnamefont {G.~M.}\
  \bibnamefont {Samarin}}, \bibinfo {author} {\bibfnamefont {G.}~\bibnamefont
  {Sarri}}, \bibinfo {author} {\bibfnamefont {D.~R.}\ \bibnamefont {Symes}},
  \bibinfo {author} {\bibfnamefont {A.~G.~R.}\ \bibnamefont {Thomas}}, \bibinfo
  {author} {\bibfnamefont {J.}~\bibnamefont {Warwick}}, \bibinfo {author}
  {\bibfnamefont {M.}~\bibnamefont {Zepf}}, \bibinfo {author} {\bibfnamefont
  {Z.}~\bibnamefont {Najmudin}}, \ and\ \bibinfo {author} {\bibfnamefont
  {S.~P.~D.}\ \bibnamefont {Mangles}},\ }\href {\doibase
  10.1103/PhysRevX.8.011020} {\bibfield  {journal} {\bibinfo  {journal}
  {Physical Review X}\ }\textbf {\bibinfo {volume} {8}},\ \bibinfo {pages}
  {011020} (\bibinfo {year} {2018})}\BibitemShut {NoStop}%
\bibitem [{\citenamefont {Poder}\ \emph
  {et~al.}(2018{\natexlab{a}})\citenamefont {Poder}, \citenamefont {Tamburini},
  \citenamefont {Sarri}, \citenamefont {Di~Piazza}, \citenamefont {Kuschel},
  \citenamefont {Baird}, \citenamefont {Behm}, \citenamefont {Bohlen},
  \citenamefont {Cole}, \citenamefont {Corvan}, \citenamefont {Duff},
  \citenamefont {Gerstmayr}, \citenamefont {Keitel}, \citenamefont
  {Krushelnick}, \citenamefont {Mangles}, \citenamefont {McKenna},
  \citenamefont {Murphy}, \citenamefont {Najmudin}, \citenamefont {Ridgers},
  \citenamefont {Samarin}, \citenamefont {Symes}, \citenamefont {Thomas},
  \citenamefont {Warwick},\ and\ \citenamefont {Zepf}}]{Poder2018PRX}%
  \BibitemOpen
  \bibfield  {author} {\bibinfo {author} {\bibfnamefont {K.}~\bibnamefont
  {Poder}}, \bibinfo {author} {\bibfnamefont {M.}~\bibnamefont {Tamburini}},
  \bibinfo {author} {\bibfnamefont {G.}~\bibnamefont {Sarri}}, \bibinfo
  {author} {\bibfnamefont {A.}~\bibnamefont {Di~Piazza}}, \bibinfo {author}
  {\bibfnamefont {S.}~\bibnamefont {Kuschel}}, \bibinfo {author} {\bibfnamefont
  {C.~D.}\ \bibnamefont {Baird}}, \bibinfo {author} {\bibfnamefont
  {K.}~\bibnamefont {Behm}}, \bibinfo {author} {\bibfnamefont {S.}~\bibnamefont
  {Bohlen}}, \bibinfo {author} {\bibfnamefont {J.~M.}\ \bibnamefont {Cole}},
  \bibinfo {author} {\bibfnamefont {D.~J.}\ \bibnamefont {Corvan}}, \bibinfo
  {author} {\bibfnamefont {M.}~\bibnamefont {Duff}}, \bibinfo {author}
  {\bibfnamefont {E.}~\bibnamefont {Gerstmayr}}, \bibinfo {author}
  {\bibfnamefont {C.~H.}\ \bibnamefont {Keitel}}, \bibinfo {author}
  {\bibfnamefont {K.}~\bibnamefont {Krushelnick}}, \bibinfo {author}
  {\bibfnamefont {S.~P.~D.}\ \bibnamefont {Mangles}}, \bibinfo {author}
  {\bibfnamefont {P.}~\bibnamefont {McKenna}}, \bibinfo {author} {\bibfnamefont
  {C.~D.}\ \bibnamefont {Murphy}}, \bibinfo {author} {\bibfnamefont
  {Z.}~\bibnamefont {Najmudin}}, \bibinfo {author} {\bibfnamefont {C.~P.}\
  \bibnamefont {Ridgers}}, \bibinfo {author} {\bibfnamefont {G.~M.}\
  \bibnamefont {Samarin}}, \bibinfo {author} {\bibfnamefont {D.~R.}\
  \bibnamefont {Symes}}, \bibinfo {author} {\bibfnamefont {A.~G.~R.}\
  \bibnamefont {Thomas}}, \bibinfo {author} {\bibfnamefont {J.}~\bibnamefont
  {Warwick}}, \ and\ \bibinfo {author} {\bibfnamefont {M.}~\bibnamefont
  {Zepf}},\ }\href {\doibase 10.1103/PhysRevX.8.031004} {\bibfield  {journal}
  {\bibinfo  {journal} {Physical Review X}\ }\textbf {\bibinfo {volume} {8}},\
  \bibinfo {pages} {031004} (\bibinfo {year} {2018}{\natexlab{a}})}\BibitemShut
  {NoStop}%
\bibitem [{\citenamefont {Blackburn}\ \emph {et~al.}(2014)\citenamefont
  {Blackburn}, \citenamefont {Ridgers}, \citenamefont {Kirk},\ and\
  \citenamefont {Bell}}]{Blackburn2014PRL}%
  \BibitemOpen
  \bibfield  {author} {\bibinfo {author} {\bibfnamefont {T.~G.}\ \bibnamefont
  {Blackburn}}, \bibinfo {author} {\bibfnamefont {C.~P.}\ \bibnamefont
  {Ridgers}}, \bibinfo {author} {\bibfnamefont {J.~G.}\ \bibnamefont {Kirk}}, \
  and\ \bibinfo {author} {\bibfnamefont {a.~R.}\ \bibnamefont {Bell}},\ }\href
  {\doibase 10.1103/PhysRevLett.112.015001} {\bibfield  {journal} {\bibinfo
  {journal} {Physical Review Letters}\ }\textbf {\bibinfo {volume} {112}},\
  \bibinfo {pages} {015001} (\bibinfo {year} {2014})}\BibitemShut {NoStop}%
\bibitem [{\citenamefont {Vranic}\ \emph {et~al.}(2016)\citenamefont {Vranic},
  \citenamefont {Grismayer}, \citenamefont {Fonseca},\ and\ \citenamefont
  {Silva}}]{Vranic2016NJP}%
  \BibitemOpen
  \bibfield  {author} {\bibinfo {author} {\bibfnamefont {M.}~\bibnamefont
  {Vranic}}, \bibinfo {author} {\bibfnamefont {T.}~\bibnamefont {Grismayer}},
  \bibinfo {author} {\bibfnamefont {R.~A.}\ \bibnamefont {Fonseca}}, \ and\
  \bibinfo {author} {\bibfnamefont {L.~O.}\ \bibnamefont {Silva}},\ }\href
  {\doibase 10.1088/1367-2630/18/7/073035} {\bibfield  {journal} {\bibinfo
  {journal} {New Journal of Physics}\ }\textbf {\bibinfo {volume} {18}},\
  \bibinfo {pages} {073035} (\bibinfo {year} {2016})}\BibitemShut {NoStop}%
\bibitem [{\citenamefont {Ridgers}\ \emph {et~al.}(2017)\citenamefont
  {Ridgers}, \citenamefont {Blackburn}, \citenamefont {Del~Sorbo},
  \citenamefont {Bradley}, \citenamefont {Slade-Lowther}, \citenamefont
  {Baird}, \citenamefont {Mangles}, \citenamefont {McKenna}, \citenamefont
  {Marklund}, \citenamefont {Murphy},\ and\ \citenamefont
  {Thomas}}]{Ridgers2017JPP}%
  \BibitemOpen
  \bibfield  {author} {\bibinfo {author} {\bibfnamefont {C.~P.}\ \bibnamefont
  {Ridgers}}, \bibinfo {author} {\bibfnamefont {T.~G.}\ \bibnamefont
  {Blackburn}}, \bibinfo {author} {\bibfnamefont {D.}~\bibnamefont
  {Del~Sorbo}}, \bibinfo {author} {\bibfnamefont {L.~E.}\ \bibnamefont
  {Bradley}}, \bibinfo {author} {\bibfnamefont {C.}~\bibnamefont
  {Slade-Lowther}}, \bibinfo {author} {\bibfnamefont {C.~D.}\ \bibnamefont
  {Baird}}, \bibinfo {author} {\bibfnamefont {S.~P.~D.}\ \bibnamefont
  {Mangles}}, \bibinfo {author} {\bibfnamefont {P.}~\bibnamefont {McKenna}},
  \bibinfo {author} {\bibfnamefont {M.}~\bibnamefont {Marklund}}, \bibinfo
  {author} {\bibfnamefont {C.~D.}\ \bibnamefont {Murphy}}, \ and\ \bibinfo
  {author} {\bibfnamefont {A.~G.~R.}\ \bibnamefont {Thomas}},\ }\href {\doibase
  10.1017/S0022377817000642} {\bibfield  {journal} {\bibinfo  {journal}
  {Journal of Plasma Physics}\ }\textbf {\bibinfo {volume} {83}},\ \bibinfo
  {pages} {715830502} (\bibinfo {year} {2017})}\BibitemShut {NoStop}%
\bibitem [{\citenamefont {Lu}\ \emph {et~al.}(2007)\citenamefont {Lu},
  \citenamefont {Tzoufras}, \citenamefont {Joshi}, \citenamefont {Tsung},
  \citenamefont {Mori}, \citenamefont {Vieira}, \citenamefont {Fonseca},\ and\
  \citenamefont {Silva}}]{Lu2007PRSTAB}%
  \BibitemOpen
  \bibfield  {author} {\bibinfo {author} {\bibfnamefont {W.}~\bibnamefont
  {Lu}}, \bibinfo {author} {\bibfnamefont {M.}~\bibnamefont {Tzoufras}},
  \bibinfo {author} {\bibfnamefont {C.}~\bibnamefont {Joshi}}, \bibinfo
  {author} {\bibfnamefont {F.}~\bibnamefont {Tsung}}, \bibinfo {author}
  {\bibfnamefont {W.}~\bibnamefont {Mori}}, \bibinfo {author} {\bibfnamefont
  {J.}~\bibnamefont {Vieira}}, \bibinfo {author} {\bibfnamefont
  {R.}~\bibnamefont {Fonseca}}, \ and\ \bibinfo {author} {\bibfnamefont
  {L.}~\bibnamefont {Silva}},\ }\href {\doibase 10.1103/PhysRevSTAB.10.061301}
  {\bibfield  {journal} {\bibinfo  {journal} {Physical Review Special Topics -
  Accelerators and Beams}\ }\textbf {\bibinfo {volume} {10}},\ \bibinfo {pages}
  {061301} (\bibinfo {year} {2007})}\BibitemShut {NoStop}%
\bibitem [{\citenamefont {Lu}\ \emph {et~al.}(2006)\citenamefont {Lu},
  \citenamefont {Huang}, \citenamefont {Zhou}, \citenamefont {Mori},\ and\
  \citenamefont {Katsouleas}}]{Lu2006PRL}%
  \BibitemOpen
  \bibfield  {author} {\bibinfo {author} {\bibfnamefont {W.}~\bibnamefont
  {Lu}}, \bibinfo {author} {\bibfnamefont {C.}~\bibnamefont {Huang}}, \bibinfo
  {author} {\bibfnamefont {M.}~\bibnamefont {Zhou}}, \bibinfo {author}
  {\bibfnamefont {W.~B.}\ \bibnamefont {Mori}}, \ and\ \bibinfo {author}
  {\bibfnamefont {T.}~\bibnamefont {Katsouleas}},\ }\href {\doibase
  10.1103/PhysRevLett.96.165002} {\bibfield  {journal} {\bibinfo  {journal}
  {Physical Review Letters}\ }\textbf {\bibinfo {volume} {96}},\ \bibinfo
  {pages} {165002} (\bibinfo {year} {2006})}\BibitemShut {NoStop}%
\bibitem [{\citenamefont {Decker}\ \emph {et~al.}(1996)\citenamefont {Decker},
  \citenamefont {Mori}, \citenamefont {Tzeng},\ and\ \citenamefont
  {Katsouleas}}]{Decker1996POP}%
  \BibitemOpen
  \bibfield  {author} {\bibinfo {author} {\bibfnamefont {C.~D.}\ \bibnamefont
  {Decker}}, \bibinfo {author} {\bibfnamefont {W.~B.}\ \bibnamefont {Mori}},
  \bibinfo {author} {\bibfnamefont {K.-C.}\ \bibnamefont {Tzeng}}, \ and\
  \bibinfo {author} {\bibfnamefont {T.}~\bibnamefont {Katsouleas}},\ }\href
  {\doibase 10.1063/1.872001} {\bibfield  {journal} {\bibinfo  {journal}
  {Physics of Plasmas}\ }\textbf {\bibinfo {volume} {3}},\ \bibinfo {pages}
  {2047} (\bibinfo {year} {1996})}\BibitemShut {NoStop}%
\bibitem [{\citenamefont {Poder}\ \emph
  {et~al.}(2018{\natexlab{b}})\citenamefont {Poder}, \citenamefont {Cole},
  \citenamefont {Wood}, \citenamefont {Lopes}, \citenamefont {Alatabi},
  \citenamefont {Foster}, \citenamefont {Kamperidis}, \citenamefont
  {Kononenko}, \citenamefont {Palmer}, \citenamefont {Rusby}, \citenamefont
  {Sahai}, \citenamefont {Sarri}, \citenamefont {Symes}, \citenamefont
  {Warwick}, \citenamefont {Mangles},\ and\ \citenamefont
  {Najmudin}}]{PoderPPCF2018}%
  \BibitemOpen
  \bibfield  {author} {\bibinfo {author} {\bibfnamefont {K.}~\bibnamefont
  {Poder}}, \bibinfo {author} {\bibfnamefont {J.~M.}\ \bibnamefont {Cole}},
  \bibinfo {author} {\bibfnamefont {J.~C.}\ \bibnamefont {Wood}}, \bibinfo
  {author} {\bibfnamefont {N.~C.}\ \bibnamefont {Lopes}}, \bibinfo {author}
  {\bibfnamefont {S.}~\bibnamefont {Alatabi}}, \bibinfo {author} {\bibfnamefont
  {P.~S.}\ \bibnamefont {Foster}}, \bibinfo {author} {\bibfnamefont
  {C.}~\bibnamefont {Kamperidis}}, \bibinfo {author} {\bibfnamefont
  {O.}~\bibnamefont {Kononenko}}, \bibinfo {author} {\bibfnamefont {C.~A.}\
  \bibnamefont {Palmer}}, \bibinfo {author} {\bibfnamefont {D.}~\bibnamefont
  {Rusby}}, \bibinfo {author} {\bibfnamefont {A.}~\bibnamefont {Sahai}},
  \bibinfo {author} {\bibfnamefont {G.}~\bibnamefont {Sarri}}, \bibinfo
  {author} {\bibfnamefont {D.~R.}\ \bibnamefont {Symes}}, \bibinfo {author}
  {\bibfnamefont {J.~R.}\ \bibnamefont {Warwick}}, \bibinfo {author}
  {\bibfnamefont {S.~P.~D.}\ \bibnamefont {Mangles}}, \ and\ \bibinfo {author}
  {\bibfnamefont {Z.}~\bibnamefont {Najmudin}},\ }\href {\doibase
  10.1088/1361-6587/aa8f0e} {\bibfield  {journal} {\bibinfo  {journal} {Plasma
  Physics and Controlled Fusion}\ }\textbf {\bibinfo {volume} {60}},\ \bibinfo
  {pages} {014022} (\bibinfo {year} {2018}{\natexlab{b}})}\BibitemShut
  {NoStop}%
\bibitem [{\citenamefont {Wang}\ \emph {et~al.}(2013)\citenamefont {Wang},
  \citenamefont {Zgadzaj}, \citenamefont {Fazel}, \citenamefont {Li},
  \citenamefont {Yi}, \citenamefont {Zhang}, \citenamefont {Henderson},
  \citenamefont {Chang}, \citenamefont {Korzekwa}, \citenamefont {Tsai},
  \citenamefont {Pai}, \citenamefont {Quevedo}, \citenamefont {Dyer},
  \citenamefont {Gaul}, \citenamefont {Martinez}, \citenamefont {Bernstein},
  \citenamefont {Borger}, \citenamefont {Spinks}, \citenamefont {Donovan},
  \citenamefont {Khudik}, \citenamefont {Shvets}, \citenamefont {Ditmire},\
  and\ \citenamefont {Downer}}]{Wang2013NC}%
  \BibitemOpen
  \bibfield  {author} {\bibinfo {author} {\bibfnamefont {X.}~\bibnamefont
  {Wang}}, \bibinfo {author} {\bibfnamefont {R.}~\bibnamefont {Zgadzaj}},
  \bibinfo {author} {\bibfnamefont {N.}~\bibnamefont {Fazel}}, \bibinfo
  {author} {\bibfnamefont {Z.}~\bibnamefont {Li}}, \bibinfo {author}
  {\bibfnamefont {S.~A.}\ \bibnamefont {Yi}}, \bibinfo {author} {\bibfnamefont
  {X.}~\bibnamefont {Zhang}}, \bibinfo {author} {\bibfnamefont
  {W.}~\bibnamefont {Henderson}}, \bibinfo {author} {\bibfnamefont {Y.-Y.}\
  \bibnamefont {Chang}}, \bibinfo {author} {\bibfnamefont {R.}~\bibnamefont
  {Korzekwa}}, \bibinfo {author} {\bibfnamefont {H.-E.}\ \bibnamefont {Tsai}},
  \bibinfo {author} {\bibfnamefont {C.-H.}\ \bibnamefont {Pai}}, \bibinfo
  {author} {\bibfnamefont {H.}~\bibnamefont {Quevedo}}, \bibinfo {author}
  {\bibfnamefont {G.}~\bibnamefont {Dyer}}, \bibinfo {author} {\bibfnamefont
  {E.}~\bibnamefont {Gaul}}, \bibinfo {author} {\bibfnamefont {M.}~\bibnamefont
  {Martinez}}, \bibinfo {author} {\bibfnamefont {A.~C.}\ \bibnamefont
  {Bernstein}}, \bibinfo {author} {\bibfnamefont {T.}~\bibnamefont {Borger}},
  \bibinfo {author} {\bibfnamefont {M.}~\bibnamefont {Spinks}}, \bibinfo
  {author} {\bibfnamefont {M.}~\bibnamefont {Donovan}}, \bibinfo {author}
  {\bibfnamefont {V.}~\bibnamefont {Khudik}}, \bibinfo {author} {\bibfnamefont
  {G.}~\bibnamefont {Shvets}}, \bibinfo {author} {\bibfnamefont
  {T.}~\bibnamefont {Ditmire}}, \ and\ \bibinfo {author} {\bibfnamefont
  {M.~C.}\ \bibnamefont {Downer}},\ }\href {\doibase 10.1038/ncomms2988}
  {\bibfield  {journal} {\bibinfo  {journal} {Nature Communications}\ }\textbf
  {\bibinfo {volume} {4}},\ \bibinfo {pages} {1988} (\bibinfo {year}
  {2013})}\BibitemShut {NoStop}%
\bibitem [{\citenamefont {Kim}\ \emph {et~al.}(2013)\citenamefont {Kim},
  \citenamefont {Pae}, \citenamefont {Cha}, \citenamefont {Kim}, \citenamefont
  {Yu}, \citenamefont {Sung}, \citenamefont {Lee}, \citenamefont {Jeong},\ and\
  \citenamefont {Lee}}]{Kim2013PRL}%
  \BibitemOpen
  \bibfield  {author} {\bibinfo {author} {\bibfnamefont {H.~T.}\ \bibnamefont
  {Kim}}, \bibinfo {author} {\bibfnamefont {K.~H.}\ \bibnamefont {Pae}},
  \bibinfo {author} {\bibfnamefont {H.~J.}\ \bibnamefont {Cha}}, \bibinfo
  {author} {\bibfnamefont {I.~J.}\ \bibnamefont {Kim}}, \bibinfo {author}
  {\bibfnamefont {T.~J.}\ \bibnamefont {Yu}}, \bibinfo {author} {\bibfnamefont
  {J.~H.}\ \bibnamefont {Sung}}, \bibinfo {author} {\bibfnamefont {S.~K.}\
  \bibnamefont {Lee}}, \bibinfo {author} {\bibfnamefont {T.~M.}\ \bibnamefont
  {Jeong}}, \ and\ \bibinfo {author} {\bibfnamefont {J.}~\bibnamefont {Lee}},\
  }\href {\doibase 10.1103/PhysRevLett.111.165002} {\bibfield  {journal}
  {\bibinfo  {journal} {Physical Review Letters}\ }\textbf {\bibinfo {volume}
  {111}},\ \bibinfo {pages} {165002} (\bibinfo {year} {2013})}\BibitemShut
  {NoStop}%
\bibitem [{\citenamefont {Butler}\ \emph {et~al.}(2002)\citenamefont {Butler},
  \citenamefont {Spence},\ and\ \citenamefont {Hooker}}]{Butler2002PRL}%
  \BibitemOpen
  \bibfield  {author} {\bibinfo {author} {\bibfnamefont {A.}~\bibnamefont
  {Butler}}, \bibinfo {author} {\bibfnamefont {D.~J.}\ \bibnamefont {Spence}},
  \ and\ \bibinfo {author} {\bibfnamefont {S.~M.}\ \bibnamefont {Hooker}},\
  }\href {\doibase 10.1103/PhysRevLett.89.185003} {\bibfield  {journal}
  {\bibinfo  {journal} {Physical Review Letters}\ }\textbf {\bibinfo {volume}
  {89}},\ \bibinfo {pages} {185003} (\bibinfo {year} {2002})}\BibitemShut
  {NoStop}%
\bibitem [{\citenamefont {Shalloo}\ \emph {et~al.}(2018)\citenamefont
  {Shalloo}, \citenamefont {Arran}, \citenamefont {Corner}, \citenamefont
  {Holloway}, \citenamefont {Jonnerby}, \citenamefont {Walczak}, \citenamefont
  {Milchberg},\ and\ \citenamefont {Hooker}}]{Shalloo2018PRE}%
  \BibitemOpen
  \bibfield  {author} {\bibinfo {author} {\bibfnamefont {R.~J.}\ \bibnamefont
  {Shalloo}}, \bibinfo {author} {\bibfnamefont {C.}~\bibnamefont {Arran}},
  \bibinfo {author} {\bibfnamefont {L.}~\bibnamefont {Corner}}, \bibinfo
  {author} {\bibfnamefont {J.}~\bibnamefont {Holloway}}, \bibinfo {author}
  {\bibfnamefont {J.}~\bibnamefont {Jonnerby}}, \bibinfo {author}
  {\bibfnamefont {R.}~\bibnamefont {Walczak}}, \bibinfo {author} {\bibfnamefont
  {H.~M.}\ \bibnamefont {Milchberg}}, \ and\ \bibinfo {author} {\bibfnamefont
  {S.~M.}\ \bibnamefont {Hooker}},\ }\href {\doibase
  10.1103/PhysRevE.97.053203} {\bibfield  {journal} {\bibinfo  {journal}
  {Physical Review E}\ }\textbf {\bibinfo {volume} {97}},\ \bibinfo {pages}
  {053203} (\bibinfo {year} {2018})}\BibitemShut {NoStop}%
\bibitem [{\citenamefont {Gonsalves}\ \emph {et~al.}(2019)\citenamefont
  {Gonsalves}, \citenamefont {Nakamura}, \citenamefont {Daniels}, \citenamefont
  {Benedetti}, \citenamefont {Pieronek}, \citenamefont {de~Raadt},
  \citenamefont {Steinke}, \citenamefont {Bin}, \citenamefont {Bulanov},
  \citenamefont {van Tilborg}, \citenamefont {Geddes}, \citenamefont
  {Schroeder}, \citenamefont {T{\'{o}}th}, \citenamefont {Esarey},
  \citenamefont {Swanson}, \citenamefont {Fan-Chiang}, \citenamefont
  {Bagdasarov}, \citenamefont {Bobrova}, \citenamefont {Gasilov}, \citenamefont
  {Korn}, \citenamefont {Sasorov},\ and\ \citenamefont
  {Leemans}}]{Gonsalves2019PRL}%
  \BibitemOpen
  \bibfield  {author} {\bibinfo {author} {\bibfnamefont {A.~J.}\ \bibnamefont
  {Gonsalves}}, \bibinfo {author} {\bibfnamefont {K.}~\bibnamefont {Nakamura}},
  \bibinfo {author} {\bibfnamefont {J.}~\bibnamefont {Daniels}}, \bibinfo
  {author} {\bibfnamefont {C.}~\bibnamefont {Benedetti}}, \bibinfo {author}
  {\bibfnamefont {C.}~\bibnamefont {Pieronek}}, \bibinfo {author}
  {\bibfnamefont {T.~C.~H.}\ \bibnamefont {de~Raadt}}, \bibinfo {author}
  {\bibfnamefont {S.}~\bibnamefont {Steinke}}, \bibinfo {author} {\bibfnamefont
  {J.~H.}\ \bibnamefont {Bin}}, \bibinfo {author} {\bibfnamefont {S.~S.}\
  \bibnamefont {Bulanov}}, \bibinfo {author} {\bibfnamefont {J.}~\bibnamefont
  {van Tilborg}}, \bibinfo {author} {\bibfnamefont {C.~G.~R.}\ \bibnamefont
  {Geddes}}, \bibinfo {author} {\bibfnamefont {C.~B.}\ \bibnamefont
  {Schroeder}}, \bibinfo {author} {\bibfnamefont {C.}~\bibnamefont
  {T{\'{o}}th}}, \bibinfo {author} {\bibfnamefont {E.}~\bibnamefont {Esarey}},
  \bibinfo {author} {\bibfnamefont {K.}~\bibnamefont {Swanson}}, \bibinfo
  {author} {\bibfnamefont {L.}~\bibnamefont {Fan-Chiang}}, \bibinfo {author}
  {\bibfnamefont {G.}~\bibnamefont {Bagdasarov}}, \bibinfo {author}
  {\bibfnamefont {N.}~\bibnamefont {Bobrova}}, \bibinfo {author} {\bibfnamefont
  {V.}~\bibnamefont {Gasilov}}, \bibinfo {author} {\bibfnamefont
  {G.}~\bibnamefont {Korn}}, \bibinfo {author} {\bibfnamefont {P.}~\bibnamefont
  {Sasorov}}, \ and\ \bibinfo {author} {\bibfnamefont {W.~P.}\ \bibnamefont
  {Leemans}},\ }\href {\doibase 10.1103/PhysRevLett.122.084801} {\bibfield
  {journal} {\bibinfo  {journal} {Physical Review Letters}\ }\textbf {\bibinfo
  {volume} {122}},\ \bibinfo {pages} {084801} (\bibinfo {year}
  {2019})}\BibitemShut {NoStop}%
\bibitem [{\citenamefont {Schroeder}\ \emph {et~al.}(2010)\citenamefont
  {Schroeder}, \citenamefont {Esarey}, \citenamefont {Geddes}, \citenamefont
  {Benedetti},\ and\ \citenamefont {Leemans}}]{Schroeder2010PRSTAB}%
  \BibitemOpen
  \bibfield  {author} {\bibinfo {author} {\bibfnamefont {C.~B.}\ \bibnamefont
  {Schroeder}}, \bibinfo {author} {\bibfnamefont {E.}~\bibnamefont {Esarey}},
  \bibinfo {author} {\bibfnamefont {C.~G.~R.}\ \bibnamefont {Geddes}}, \bibinfo
  {author} {\bibfnamefont {C.}~\bibnamefont {Benedetti}}, \ and\ \bibinfo
  {author} {\bibfnamefont {W.~P.}\ \bibnamefont {Leemans}},\ }\href {\doibase
  10.1103/PhysRevSTAB.13.101301} {\bibfield  {journal} {\bibinfo  {journal}
  {Physical Review Special Topics - Accelerators and Beams}\ }\textbf {\bibinfo
  {volume} {13}},\ \bibinfo {pages} {101301} (\bibinfo {year}
  {2010})}\BibitemShut {NoStop}%
\bibitem [{\citenamefont {Berestetskii}\ \emph {et~al.}(1982)\citenamefont
  {Berestetskii}, \citenamefont {Lifshitz},\ and\ \citenamefont
  {Pitaevskii}}]{Berestetskii1982quantum}%
  \BibitemOpen
  \bibfield  {author} {\bibinfo {author} {\bibfnamefont {V.~B.}\ \bibnamefont
  {Berestetskii}}, \bibinfo {author} {\bibfnamefont {E.~M.}\ \bibnamefont
  {Lifshitz}}, \ and\ \bibinfo {author} {\bibfnamefont {L.~P.}\ \bibnamefont
  {Pitaevskii}},\ }\href@noop {} {\emph {\bibinfo {title} {{Quantum
  electrodynamics}}}},\ Vol.~\bibinfo {volume} {4}\ (\bibinfo  {publisher}
  {Butterworth-Heinemann},\ \bibinfo {year} {1982})\BibitemShut {NoStop}%
\bibitem [{\citenamefont {Seipt}\ \emph {et~al.}(2015)\citenamefont {Seipt},
  \citenamefont {Rykovanov}, \citenamefont {Surzhykov},\ and\ \citenamefont
  {Fritzsche}}]{Seipt2015PRA}%
  \BibitemOpen
  \bibfield  {author} {\bibinfo {author} {\bibfnamefont {D.}~\bibnamefont
  {Seipt}}, \bibinfo {author} {\bibfnamefont {S.~G.}\ \bibnamefont
  {Rykovanov}}, \bibinfo {author} {\bibfnamefont {A.}~\bibnamefont
  {Surzhykov}}, \ and\ \bibinfo {author} {\bibfnamefont {S.}~\bibnamefont
  {Fritzsche}},\ }\href {\doibase 10.1103/PhysRevA.91.033402} {\bibfield
  {journal} {\bibinfo  {journal} {Physical Review A - Atomic, Molecular, and
  Optical Physics}\ }\textbf {\bibinfo {volume} {91}},\ \bibinfo {pages}
  {033402} (\bibinfo {year} {2015})}\BibitemShut {NoStop}%
\bibitem [{\citenamefont {Seipt}\ and\ \citenamefont
  {K{\"{a}}mpfer}(2011)}]{Seipt2011PRA}%
  \BibitemOpen
  \bibfield  {author} {\bibinfo {author} {\bibfnamefont {D.}~\bibnamefont
  {Seipt}}\ and\ \bibinfo {author} {\bibfnamefont {B.}~\bibnamefont
  {K{\"{a}}mpfer}},\ }\href {\doibase 10.1103/PhysRevA.83.022101} {\bibfield
  {journal} {\bibinfo  {journal} {Physical Review A}\ }\textbf {\bibinfo
  {volume} {83}},\ \bibinfo {pages} {022101} (\bibinfo {year}
  {2011})}\BibitemShut {NoStop}%
\bibitem [{\citenamefont {Seipt}(2020)}]{Seipt2020PC}%
  \BibitemOpen
  \bibfield  {author} {\bibinfo {author} {\bibfnamefont {D.}~\bibnamefont
  {Seipt}},\ }\href@noop {} {\bibfield  {journal} {\bibinfo  {journal} {Private
  Communication}\ } (\bibinfo {year} {2020})}\BibitemShut {NoStop}%
\bibitem [{\citenamefont {Floettmann}(2003)}]{Floettmann2003}%
  \BibitemOpen
  \bibfield  {author} {\bibinfo {author} {\bibfnamefont {K.}~\bibnamefont
  {Floettmann}},\ }\href {\doibase 10.1103/PhysRevSTAB.6.034202} {\bibfield
  {journal} {\bibinfo  {journal} {Physical Review Special Topics - Accelerators
  and Beams}\ }\textbf {\bibinfo {volume} {6}},\ \bibinfo {pages} {80}
  (\bibinfo {year} {2003})}\BibitemShut {NoStop}%
\bibitem [{\citenamefont {Mangles}\ \emph {et~al.}(2006)\citenamefont
  {Mangles}, \citenamefont {Thomas}, \citenamefont {Kaluza}, \citenamefont
  {Lundh}, \citenamefont {Lindau}, \citenamefont {Persson}, \citenamefont
  {Tsung}, \citenamefont {Najmudin}, \citenamefont {Mori}, \citenamefont
  {Wahlstrom},\ and\ \citenamefont {Krushelnick}}]{Mangles2006PRL}%
  \BibitemOpen
  \bibfield  {author} {\bibinfo {author} {\bibfnamefont {S.~P.~D.}\
  \bibnamefont {Mangles}}, \bibinfo {author} {\bibfnamefont {A.~G.~R.}\
  \bibnamefont {Thomas}}, \bibinfo {author} {\bibfnamefont {M.~C.}\
  \bibnamefont {Kaluza}}, \bibinfo {author} {\bibfnamefont {O.}~\bibnamefont
  {Lundh}}, \bibinfo {author} {\bibfnamefont {F.}~\bibnamefont {Lindau}},
  \bibinfo {author} {\bibfnamefont {A.}~\bibnamefont {Persson}}, \bibinfo
  {author} {\bibfnamefont {F.~S.}\ \bibnamefont {Tsung}}, \bibinfo {author}
  {\bibfnamefont {Z.}~\bibnamefont {Najmudin}}, \bibinfo {author}
  {\bibfnamefont {W.~B.}\ \bibnamefont {Mori}}, \bibinfo {author}
  {\bibfnamefont {C.~G.}\ \bibnamefont {Wahlstrom}}, \ and\ \bibinfo {author}
  {\bibfnamefont {K.}~\bibnamefont {Krushelnick}},\ }\href {\doibase 10.1103/PhysRevLett.96.215001} {\bibfield
  {journal} {\bibinfo  {journal} {Physical Review Letters}\ }\textbf {\bibinfo
  {volume} {96}},\ \bibinfo {pages} {215001} (\bibinfo {year}
  {2006})}\BibitemShut {NoStop}%
\bibitem [{\citenamefont {Hooker}\ \emph {et~al.}(2014)\citenamefont {Hooker},
  \citenamefont {Bartolini}, \citenamefont {Mangles}, \citenamefont
  {T{\"{u}}nnermann}, \citenamefont {Corner}, \citenamefont {Limpert},
  \citenamefont {Seryi},\ and\ \citenamefont {Walczak}}]{Hooker2014JPB}%
  \BibitemOpen
  \bibfield  {author} {\bibinfo {author} {\bibfnamefont {S.~M.}\ \bibnamefont
  {Hooker}}, \bibinfo {author} {\bibfnamefont {R.}~\bibnamefont {Bartolini}},
  \bibinfo {author} {\bibfnamefont {S.~P.~D.}\ \bibnamefont {Mangles}},
  \bibinfo {author} {\bibfnamefont {A.}~\bibnamefont {T{\"{u}}nnermann}},
  \bibinfo {author} {\bibfnamefont {L.}~\bibnamefont {Corner}}, \bibinfo
  {author} {\bibfnamefont {J.}~\bibnamefont {Limpert}}, \bibinfo {author}
  {\bibfnamefont {A.}~\bibnamefont {Seryi}}, \ and\ \bibinfo {author}
  {\bibfnamefont {R.}~\bibnamefont {Walczak}},\ }\href {\doibase
  10.1088/0953-4075/47/23/234003} {\bibfield  {journal} {\bibinfo  {journal}
  {Journal of Physics B: Atomic, Molecular and Optical Physics}\ }\textbf
  {\bibinfo {volume} {47}},\ \bibinfo {pages} {234003} (\bibinfo {year}
  {2014})}\BibitemShut {NoStop}%
\bibitem [{\citenamefont {Luo}\ \emph {et~al.}(2018)\citenamefont {Luo},
  \citenamefont {Chen}, \citenamefont {Wu}, \citenamefont {Weng}, \citenamefont
  {Sheng}, \citenamefont {Schroeder}, \citenamefont {Jaroszynski},
  \citenamefont {Esarey}, \citenamefont {Leemans}, \citenamefont {Mori},\ and\
  \citenamefont {Zhang}}]{Luo2018PRL}%
  \BibitemOpen
  \bibfield  {author} {\bibinfo {author} {\bibfnamefont {J.}~\bibnamefont
  {Luo}}, \bibinfo {author} {\bibfnamefont {M.}~\bibnamefont {Chen}}, \bibinfo
  {author} {\bibfnamefont {W.~Y.}\ \bibnamefont {Wu}}, \bibinfo {author}
  {\bibfnamefont {S.~M.}\ \bibnamefont {Weng}}, \bibinfo {author}
  {\bibfnamefont {Z.~M.}\ \bibnamefont {Sheng}}, \bibinfo {author}
  {\bibfnamefont {C.~B.}\ \bibnamefont {Schroeder}}, \bibinfo {author}
  {\bibfnamefont {D.~A.}\ \bibnamefont {Jaroszynski}}, \bibinfo {author}
  {\bibfnamefont {E.}~\bibnamefont {Esarey}}, \bibinfo {author} {\bibfnamefont
  {W.~P.}\ \bibnamefont {Leemans}}, \bibinfo {author} {\bibfnamefont {W.~B.}\
  \bibnamefont {Mori}}, \ and\ \bibinfo {author} {\bibfnamefont
  {J.}~\bibnamefont {Zhang}},\ }\href {\doibase 10.1103/PhysRevLett.120.154801}
  {\bibfield  {journal} {\bibinfo  {journal} {Physical Review Letters}\
  }\textbf {\bibinfo {volume} {120}},\ \bibinfo {pages} {154801} (\bibinfo
  {year} {2018})}\BibitemShut {NoStop}%
\bibitem [{\citenamefont {Walker}\ \emph {et~al.}(2017)\citenamefont {Walker},
  \citenamefont {Alesini}, \citenamefont {Alexandrova}, \citenamefont {Anania},
  \citenamefont {Andreev}, \citenamefont {Andriyash}, \citenamefont
  {Aschikhin}, \citenamefont {Assmann}, \citenamefont {Audet}, \citenamefont
  {Bacci}, \citenamefont {Barna}, \citenamefont {Beaton}, \citenamefont {Beck},
  \citenamefont {Beluze}, \citenamefont {Bernhard}, \citenamefont {Bielawski},
  \citenamefont {Bisesto}, \citenamefont {Boedewadt}, \citenamefont {Brandi},
  \citenamefont {Bringer}, \citenamefont {Brinkmann}, \citenamefont
  {Br{\"{u}}ndermann}, \citenamefont {B{\"{u}}scher}, \citenamefont {Bussmann},
  \citenamefont {Bussolino}, \citenamefont {Chance}, \citenamefont
  {Chanteloup}, \citenamefont {Chen}, \citenamefont {Chiadroni}, \citenamefont
  {Cianchi}, \citenamefont {Clarke}, \citenamefont {Cole}, \citenamefont
  {Couprie}, \citenamefont {Croia}, \citenamefont {Cros}, \citenamefont {Dale},
  \citenamefont {Dattoli}, \citenamefont {Delerue}, \citenamefont
  {Delferriere}, \citenamefont {Delinikolas}, \citenamefont {Dias},
  \citenamefont {Dorda}, \citenamefont {Ertel}, \citenamefont {Ferran~Pousa},
  \citenamefont {Ferrario}, \citenamefont {Filippi}, \citenamefont {Fils},
  \citenamefont {Fiorito}, \citenamefont {Fonseca}, \citenamefont {Galimberti},
  \citenamefont {Gallo}, \citenamefont {Garzella}, \citenamefont {Gastinel},
  \citenamefont {Giove}, \citenamefont {Giribono}, \citenamefont {Gizzi},
  \citenamefont {Gr{\"{u}}ner}, \citenamefont {Habib}, \citenamefont {Haefner},
  \citenamefont {Heinemann}, \citenamefont {Hidding}, \citenamefont {Holzer},
  \citenamefont {Hooker}, \citenamefont {Hosokai}, \citenamefont {Irman},
  \citenamefont {Jaroszynski}, \citenamefont {Jaster-Merz}, \citenamefont
  {Joshi}, \citenamefont {Kaluza}, \citenamefont {Kando}, \citenamefont
  {Karger}, \citenamefont {Karsch}, \citenamefont {Khazanov}, \citenamefont
  {Khikhlukha}, \citenamefont {Knetsch}, \citenamefont {Kocon}, \citenamefont
  {Koester}, \citenamefont {Kononenko}, \citenamefont {Korn}, \citenamefont
  {Kostyukov}, \citenamefont {Labate}, \citenamefont {Lechner}, \citenamefont
  {Leemans}, \citenamefont {Lehrach}, \citenamefont {Li}, \citenamefont {Li},
  \citenamefont {Libov}, \citenamefont {Lifschitz}, \citenamefont {Litvinenko},
  \citenamefont {Lu}, \citenamefont {Maier}, \citenamefont {Malka},
  \citenamefont {Manahan}, \citenamefont {Mangles}, \citenamefont {Marchetti},
  \citenamefont {Marocchino}, \citenamefont {Martinez De La~Ossa},
  \citenamefont {Martins}, \citenamefont {Massimo}, \citenamefont {Mathieu},
  \citenamefont {Maynard}, \citenamefont {Mehrling}, \citenamefont
  {Molodozhentsev}, \citenamefont {Mosnier}, \citenamefont {Mostacci},
  \citenamefont {Mueller}, \citenamefont {Najmudin}, \citenamefont {Nghiem},
  \citenamefont {Nguyen}, \citenamefont {Niknejadi}, \citenamefont {Osterhoff},
  \citenamefont {Papadopoulos}, \citenamefont {Patrizi}, \citenamefont
  {Pattathil}, \citenamefont {Petrillo}, \citenamefont {Pocsai}, \citenamefont
  {Poder}, \citenamefont {Pompili}, \citenamefont {Pribyl}, \citenamefont
  {Pugacheva}, \citenamefont {Romeo}, \citenamefont {Rossi}, \citenamefont
  {Roussel}, \citenamefont {Sahai}, \citenamefont {Scherkl}, \citenamefont
  {Schramm}, \citenamefont {Schroeder}, \citenamefont {Schwindling},
  \citenamefont {Scifo}, \citenamefont {Serafini}, \citenamefont {Sheng},
  \citenamefont {Silva}, \citenamefont {Silva}, \citenamefont {Simon},
  \citenamefont {Sinha}, \citenamefont {Specka}, \citenamefont {Streeter},
  \citenamefont {Svystun}, \citenamefont {Symes}, \citenamefont {Szwaj},
  \citenamefont {Tauscher}, \citenamefont {Thomas}, \citenamefont {Thompson},
  \citenamefont {Toci}, \citenamefont {Tomassini}, \citenamefont {Vaccarezza},
  \citenamefont {Vannini}, \citenamefont {Vieira}, \citenamefont {Villa},
  \citenamefont {Wahlstr{\"{o}}m}, \citenamefont {Walczak}, \citenamefont
  {Weikum}, \citenamefont {Welsch}, \citenamefont {Wiemann}, \citenamefont
  {Wolfenden}, \citenamefont {Xia}, \citenamefont {Yabashi}, \citenamefont
  {Yu}, \citenamefont {Zhu},\ and\ \citenamefont {Zigler}}]{Walker2017JP}%
  \BibitemOpen
  \bibfield  {author} {\bibinfo {author} {\bibfnamefont {P.}~\bibnamefont
  {Walker, \emph{et al.}}} }\href {\doibase
  10.1088/1742-6596/874/1/012029} {\bibfield  {journal} {\bibinfo  {journal}
  {Journal of Physics: Conference Series}\ }\textbf {\bibinfo {volume} {874}}
  (\bibinfo {year} {2017}),\ 10.1088/1742-6596/874/1/012029}\BibitemShut
  {NoStop}%
\end{thebibliography}
\end{document}